\documentclass[aps,prb,reprint,showpacs,superscriptaddress,floatfix]{revtex4-2}
\usepackage[pdftex]{graphicx}
\usepackage{bm}
\usepackage[usenames, dvipsnames]{color}
\usepackage{comment}

\usepackage{xr-hyper}

\usepackage{hyperref}
\hypersetup{
     colorlinks   = true,
     citecolor    = blue,
     linkcolor  =black,
     urlcolor   =black}

\usepackage[T1]{fontenc}

\usepackage{colonequals}
\usepackage{amsfonts}
\usepackage{amssymb}
\usepackage{amsmath}
\usepackage{amsthm}
\usepackage{bbm}
\usepackage{multirow}

\usepackage{wasysym}

\newcommand{\e}{\mathrm{e}}

\def\talpha{\bar{\alpha}}
\def\tbeta{\bar{\beta}}

\def\bra#1{\langle#1|}
\def\ket#1{|#1\rangle}
\def\braket#1#2{\langle#1|#2\rangle}

\def\cexp#1{\langle#1\rangle}

\begin{document}
\title{Universal non-power law scaling from a chaotic RG in the Harper-Hofstadter model}
\author{Luke Yeo}
\affiliation{Department of Physics, University of Illinois at Urbana-Champaign, Urbana, Illinois, USA}
\affiliation{Office of Gas and Electricity Markets (Ofgem), Glasgow G1 1LH,
  United Kingdom}

\author{Philip J. D. Crowley}
\email{philip.jd.crowley@gmail.com}
\affiliation{Department of Physics, Massachusetts Institute of Technology, Cambridge, Massachusetts 02139, USA}
\affiliation{Department of Physics and Astronomy, Michigan State University, East Lansing, Michigan 48824, USA}

\date{\today}

\begin{abstract}
Previous studies of incommensurate systems concluded that their critical scaling is sensitively dependent on the irrational, $\alpha$, which determines the incommensuration. Contrary to this belief, in the canonical Harper-Hofstadter model, we show there is universal $\alpha$-independent scaling for almost all $\alpha$. This critical scaling is characterized by non-power-law time-length scaling $t \sim r^{\zeta \log \log r}$. We demonstrate this in the superfluid fraction of a Bose gas, and the specific heat of a Fermi gas. This scaling is generic of a broad class of generalized Harper-Hofstadter models. The $\alpha$-independent scaling emerges as the number theoretic properties of almost all irrational numbers are statistically identical (\emph{\`{a} la} Gauss-Kuzmin statistics). Consequently, we conjecture similar $\alpha$-independent scaling applies in incommensurate models more generally.
\end{abstract}

\maketitle

An incommensurate system is characterized by an irrational ratio, $\alpha$, of two microscopic scales, often the lattice length to the wavelength of a modulating potential, or magnetic flux per unit cell. These systems are realized in synthetic electronic materials~\cite{dean2013hofstadter,hunt2013massive,ponomarenko2013cloning,andrei2020graphene,carr2017twistronics,uri2023superconductivity} and cold atoms experiments~\cite{roati2008anderson,deissler2010delocalization,aidelsburger2013realization,schreiber2015observation,bordia2017probing}.

Under sufficient coarse graining, a generic critical system approaches a scaling limit in which its properties are invariant under rescalings of space and time $(r,t) \to (r / \lambda, t/ \tau )$, related by the dynamical exponent $z$, $\tau \sim \lambda^z$~\cite{goldenfeld2018lectures}. This scaling limit is revealed by renormalization group (RG) analysis~\cite{cardy1996scaling}. Under the RG the microscopic parameters $\vec{g}$ flow $\partial_s \vec{g} = \beta(\vec{g})$, where $s = \log \lambda$. The critical scaling limit is obtained from the flow's fixed point $\beta(\vec{g}_{\star}) = 0$.

Critical incommensurate systems do not have such a scaling limit. Instead, coarse graining reveals a hierarchy of ever longer ``microscopic'' length-scales set by the continued fraction expansion (CFE) of $\alpha$~\cite{simon1982almost,damanik2009spectrum}. Nevertheless, for special $\alpha$, there exist RG fixed points, of a \emph{discrete} scale transformation $(\alpha_\star , \vec{g}_\star) = B(\alpha_\star,\vec{g}_\star)$~\cite{kohmoto1983localization,ostlund1984renormalization,wurtz1988renormalization,levitov1989renormalization,hermisson1997aperiodic,vieira2005aperiodic,thiem2015origin}, where we include $\alpha$ as a microscopic parameter. The discrete transformation $B$ rescales space by $\lambda$, and time by $\tau \sim \lambda^z$ again yielding power-law dynamical scaling. However, generic incommensurate systems do not exhibit scale invariance. Instead, $\alpha$ changes with each RG step $(\alpha' , \vec{g}_\star) = B(\alpha,\vec{g}_\star)$, and so do the length rescaling and dynamical exponent. Thus, after $N$ RG steps, length is rescaled by $\lambda = \lambda_1 \lambda_2 \cdots \lambda_N$, and time by $\tau = \lambda_1^{z_1}\lambda_2^{z_2}\cdots \lambda_N^{z_N} $~\cite{suslov1982localization,wilkinson1984critical,wilkinson1987exact,szabo2018non}. Moreover, the sequence of $(\lambda_n,z_n)$ depends sensitively on the CFE of $\alpha$, appearing to preclude universal dynamical scaling~\cite{szabo2018non}. 

Though the CFEs for distinct $\alpha$ are themselves distinct, for almost all $\alpha$ they have identical statistical properties, set by Gauss-Kuzmin statistics~\cite{khinchin1963continued}. Adapting tools from Gauss-Kuzmin statistics, we show that in fact there \emph{is} $\alpha$-independent universal scaling. Specifically, in the critical Harper-Hofstadter (HH) model, for almost all $\alpha$, we show the distribution of space-time rescalings (i.e. of the $(\lambda_n,z_n)$ over $n$) is $\alpha$-independent. Naively, this implies power-law scaling, with dynamical exponent
\begin{equation}
    z = \lim_{\lambda \to \infty} \frac{\log \tau}{\log \lambda} = \lim_{N \to \infty} \frac{\sum_{n=1}^N z_n \log \lambda_n }{\sum_{n=1}^N \log \lambda_n}.
    \label{eq:1}
\end{equation}
However this is incorrect: the limit in Eq.~\eqref{eq:1} does not exist due to rare RG steps with large $z_n$. These steps occur when the renormalized $\alpha_n$ is anomalously close to rational, i.e. the renormalized model is almost commensurate. Instead, we obtain the scaling
\begin{equation}
    \tau \sim \lambda^{z(\zeta,\lambda)} \qquad z(\zeta,\lambda) := \zeta \log |\log \lambda | + O(1)
    \label{eq:dyn_scaling}
\end{equation}
where $\zeta$ is an $\alpha$-independent constant.

This RG flow constitutes a novel type of scaling. The flow does not approach a fixed point $(\alpha_\star,g_\star)$. Instead, the flow ergodically explores a region of parameter space, and the asymptotic scaling~\eqref{eq:dyn_scaling} is set by the steady state distribution of the flow $f(\alpha,g)$.

\paragraph*{\textbf{Model}:} Consider a free electron in a magnetic field, and a 2D periodic potential $V$. At strong field we may project into a single Landau level, yielding an HH model~\footnote{Eq.~\eqref{eq:H1} maps to Hofstadter's square lattice problem upon exchanging position and momenta $(\hat{x},\hat{y}) \to (\hat{p}_x,\hat{p}_y)$~\cite{zilberman1956zh,thouless1983wavefunction}}
\begin{equation}
    H = V(\hat x,\hat y) :=  - 2 V_x \cos \hat x - 2 V_y \cos \hat y, \quad [\hat x,\hat y] = 2 \pi i \alpha,
    \label{eq:H1}
\end{equation}
where $2 \pi/ \alpha$ is the flux per unit cell~\cite{rauh1974bloch,zilberman1956zh,thouless1983wavefunction,paul2022moir}. Without loss of generality, we set $\alpha \in [0,1)$ and $V_x,V_y >0$.

$H$ may be recast as a quasiperiodically modulated, 1D tight-binding model. Written in the $x$ basis $H$ couples only the points $x_n \in x_0 + 2 \pi \alpha \mathbb{N}$. Thus $H$ decomposes into a sum of decoupled sectors $H = \int_0^{2\pi\alpha} d x_0 H_\mathrm{AA}(x_0)$ where $H_\mathrm{AA}$ is  the Aubry-Andre (AA) model~\cite{aubry1980analyticity,azbel1964zh,azbel1979quantum}
\begin{equation}
    H_\mathrm{AA} = - \sum_{n \in \mathbb{Z}}   \Big[ V_y \left(  \ket{n+1}\bra{n} +\mathrm{h.c.} \right) + 2 V_x \cos x_n \ket{n}\bra{n} \Big],
    \label{eq:AA}
\end{equation}
where $\hat{x}\ket{n} = x_n\ket{n}$ and $x_n =x_0 + 2 \pi \alpha n$.

The HH model has two phases. For $V_x > V_y$, the eigenmodes of $H$ are localized (extended) in the $y$ ($x$) direction, and \emph{vice versa} for $V_x < V_y$~\cite{jitomirskaya1999metal}. $H_\mathrm{AA}$ has a corresponding metal-insulator transition, with ballistically spreading (localized) modes for $V_x > V_y, \, (V_x < V_y)$.

At the critical point $V_x = V_y$, the eigenmodes of $H$ are critically delocalized in both directions, and wavepackets spread sub-ballistically~\cite{hiramoto1988dynamicsII,ketzmerick1997determines}. Close to $V_x = V_y$, critical behaviour is found on scales below the correlation length
\begin{equation}
    \xi = |\delta|^{- \nu},\qquad \text{with} \qquad \delta = \log (V_x/V_y), \quad \nu = 1.
    \label{eq:xi}
\end{equation}

\paragraph*{\textbf{Bandstructure for rational $\alpha$:}}

\begin{figure}[t!]
    \centering
    \includegraphics[width=0.9\linewidth]{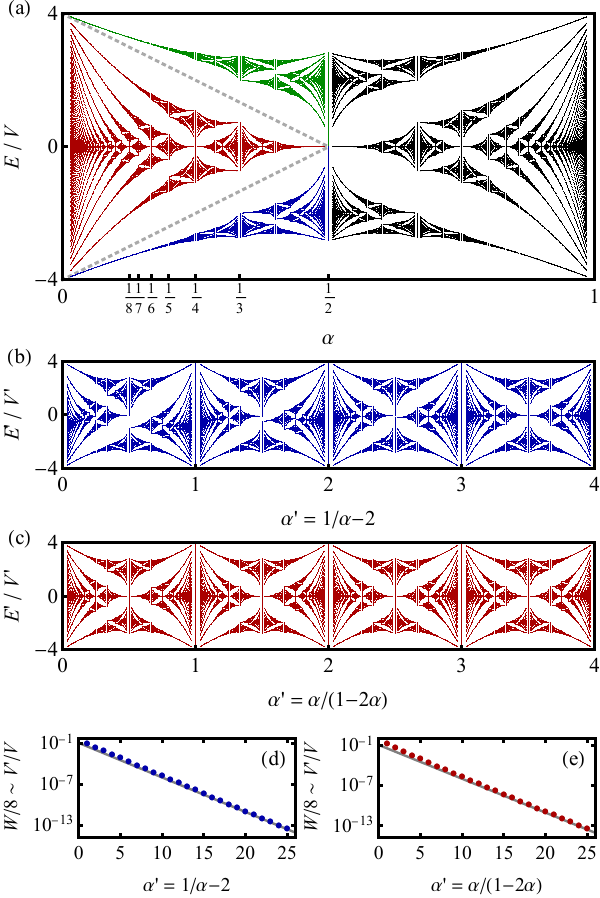}
    \caption{
    \emph{Hofstadter's butterfly}: The spectrum of~\eqref{eq:H1} is plotted (a) for all $\alpha = p/q$ with $q \leq 100$. For $\alpha < 1/2$ the upper (green), central (red), and lower (blue) Hofstadter bands are shaded. The bands are demarcated by the lines $E/V = \pm (4 - 8 \alpha)$ (grey). We re-plot the lower (b) and central (c) with appropriate shifting/rescaling of energy. In (d,e) the energetic rescaling used to create (b,c) is plotted, and compared with the analytic form~\eqref{eq:V_scaling} (grey line).
    }
    \label{Fig:spectrum}
\end{figure}

The spectrum of $H$ consists of the eigenenergies of $H_\mathrm{AA}$ calculated for all $x_0$. Plotted versus $\alpha$, the spectrum forms \emph{Hofstadter's butterfly} (Fig.~\ref{Fig:spectrum})~\cite{hofstadter1976energy,satija2016butterfly}. We discuss the properties of this fractal spectrum underpinning our RG analysis.

For irrational $\alpha$, the spectrum forms a Cantor set~\cite{avila2006solving,damanik2009spectrum}. However, for rational $\alpha$ ($\alpha = p/q$ for coprime $p,q$) $H_\mathrm{AA}$ is $q$-site periodic, and the spectrum consists of bands $j = 1 \ldots q$ with energies $E_j(x_0,y_0)$. Here $y_0$ (dual to $x_0$) is the crystal momentum. $E_j(x_0,y_0)$ is periodic with a ``Brillouin zone'' $-\pi/q < x_0,y_0 \leq \pi/q$.

Each band $E_j$ has a simple dependency on $(x_0,y_0)$, resembling the original potential $V(x,y)$~\cite{chambers1965linear,bellissard1982cantor,thouless1983bandwidths,wilkinson1984critical,wilkinson1987exact,thouless1990scaling,thouless1990scaling,last1994zero,last1992sum} 
\begin{equation}
    E_j(x_0,y_0) =  E_{j}^* - 2 V_x' \cos q x_0 - 2 V_y' \cos q y_0 + O(W_j^2/\Delta_j)
    \label{eq:Ea}
\end{equation}
(see App.~\ref{app:chambers}) where $|V_x'/V_y'| = |V_x/V_y|^q$, $\Delta_j$ is the gap to the next band and $W_j$ is the bandwidth. This self-similarity underpins the RG.

A limit in which the corrections to~\eqref{eq:Ea} vanish is the critical point $V = V_x = V_y$, for $p=1$, and $q$ large. We find this case dominates the scaling behaviour of the curvature of the lowest band. In this limit the gaps (bandwidths) are power-law (exponentially) small~\cite{wilkinson1984critical,wilkinson1987exact}
\begin{equation}
    \Delta_j \sim V/[q \, \varrho(E_{j}^*/V)],
    \quad 
    \log W_j \sim -q \, \ell(E_{j}^*/V).
    \label{eq:Wgap_size}
\end{equation}
Here $\varrho(z)$ and $\ell(z)$ are positive, even, $q$-independent functions, finite and smooth for $z \neq 0$~\footnote{at $z=0$, $\ell(0) = 0$ and $\varrho(z)$ diverges logarithmically} (see App.~\ref{app:89}).

Eq.~\eqref{eq:Ea} can be understood as a rescaled HH model. Specifically, if we define rescaled lengths $\hat{x}' = q x_0$, $\hat{y}' = q y_0$, then $H' = E_j(\hat{x}',\hat{y}')$ is the renormalized Hamiltonian obtained by projecting into the $j$th band. $H'$ forms a copy of the original model~\eqref{eq:H1} for the trivial case of $\alpha' = 0$, with time rescaled by $\tau = V/V' = 8V/W_j$, and length by $\lambda = q$. 

By these arguments, the bandstructure takes a simple form for $\alpha = 1/q$ as $q \to \infty$. Similar arguments apply for $\alpha$ approaching all other rationals. E.g. consider $q = 2 p +1$. At large $q$, $\alpha \to 1/2$. In this limit too, the band-gaps  decay as power-law in $q$, and the bandwidths exponentially in $q$ (see App.~\ref{app:WKB}). Thus, here too, the corrections to~\eqref{eq:Ea} are vanishing in $q$. We find this case dominates scaling behaviour of the heat capacity of a half-filled Fermi gas.

\paragraph*{\textbf{RG transformation:} } As above, for rational $\alpha$, projecting into a single band of the HH model yields a rescaled HH model with $\alpha' = 0$. This is a special case of an RG transformation which applies for all $\alpha$, and generically yields renormalized $\alpha' \neq 0$.

The RG transformation consists of a projection into a Hofstadter band. Hofstadter bands generalize the notion of bands to irrational $\alpha$. The spectrum consists of three Hofstadter bands Fig.~\ref{Fig:spectrum}a. As in~\eqref{eq:Ea}, projecting into a Hofstadter band yields an HH model with renormalized parameters, plus $O(W_j^2/\Delta_j)$ corrections. 

We now state the RG step for the critical HH model. The RG step depends on whether one projects an outer band, or the central band. For an outer band, we obtain lattice length rescaling
\begin{subequations}
\begin{equation}
    a \to a' = a \lambda(\talpha),
    \label{eq:space_scaling}
\end{equation}
with scale factor $\lambda(z)=1/z$. We use the symmetry of the spectrum to define the RG step in terms of the flux density parameter $\talpha : = \min(\alpha,1-\alpha)$, which renormalizes
\begin{equation}
    \talpha \to \talpha' = B(\talpha) : = \min[ b(\talpha), 1- b(\talpha)] 
    \label{eq:beta_scaling}
\end{equation}
where $b(z) := s(z) - \lfloor s(z) \rfloor$ with $s(z) = 1/z$~\footnote{In this case, $b(z)$ is the Gauss map.}; and an energy rescaling, given at large $s(\talpha)$ by
\begin{equation}
    \log V \to \log V' \sim \log V - 4 G s(\talpha)/\pi
    \label{eq:V_scaling}
\end{equation}\label{eq:RG}\end{subequations}
where $G = 0.9159\ldots$ is Catalan's constant, and $\sim$ denotes asymptotic equality. Projecting into the central band yields the same RG rules~\eqref{eq:RG} with $\lambda(z)= 1/(1-2z)$, $s(z) =z/(1-2z)$. 

The RG transformation~\eqref{eq:RG} follows from the RG analyses of Refs.~\cite{suslov1982localization,wilkinson1984critical,wilkinson1987exact,szabo2018non} and the Hofstadter rules~\cite{hofstadter1976energy,rudinger1997hofstadter}.

The length rescaling~\eqref{eq:space_scaling} follows from the fraction of states in each Hofstadter band. At a given $\talpha$, a fraction $\talpha$ of the spectrum is in each outer band, and the remaining $1-2\talpha$ is in the central band~\cite{wannier1978result}. The rescaling~\eqref{eq:space_scaling} ensures a fixed spatial density of sites under the RG. 

The renormalization of $\talpha$~\eqref{eq:beta_scaling} follows from the Hofstadter rules~\cite{hofstadter1976energy,rudinger1997hofstadter}. Specifically, for $\alpha \in [0,1/2]$, either (i) projecting into an outer band, and sending $\alpha \to \alpha' = 1/\alpha - 2$, or (ii) projecting into the central band, and sending $\alpha \to \alpha' = \alpha/(1-2\alpha)$, followed in either case by $\alpha'$-dependent energy shifts/rescaling, yields a copy of the Hofstadter spectrum up to corrections that are small in $\alpha'$. In Figs.~\ref{Fig:spectrum}b,~\ref{Fig:spectrum}c the lower (blue) and central (red) Hofstadter bands are plotted versus $\alpha'$ with appropriate $\alpha'$-dependent energy shifts/rescaling to match the envelope of the bare spectrum. Eq.~\eqref{eq:beta_scaling} simply recasts these relations in terms of $\talpha$.

The energy rescaling~\eqref{eq:V_scaling} follows from WKB analysis (see App.~\ref{app:WKB} and Refs.~\cite{szabo2018non,wilkinson1984critical,wilkinson1987exact,han1994critical,landau1981quantum,helffersjostrand1988,*helffersjostrand1990,*helffersjostrand1989}), and is numerically verified in Figs.~\ref{Fig:spectrum}d,~\ref{Fig:spectrum}e.

\paragraph*{\textbf{Curvature of the lowest band}:} Consider the curvature of the minimum of the lowest band of $H_\mathrm{AA}$
\begin{equation}
    \Gamma = \left. \frac{d^2 E_0}{d y_0^2} \right|_{y_0 = y_{0,\min}}.
    \label{eq:Gam}
\end{equation}
This quantity measures the energetic change of low-energy particles upon passing a current, and, in the many-boson case, determines the superfluid fraction~\cite{schultka1994finite,lieb2002superfluidity,roth2003phase,cestari2010finite,szabo2018non}. $\Gamma$ is non-zero only in the delocalized phase ($\delta < 0$), the approach to zero is set by the critical scaling
\begin{equation}
    \Gamma \sim \delta^{\nu[  z(\zeta,\delta)-2]}, \qquad \zeta = 48 G / \pi^3.
    \label{eq:gamma_scaling0}
\end{equation}

We obtain~\eqref{eq:gamma_scaling0} from the finite-$q$ scaling of $\Gamma$, found by terminating the RG after finitely many steps. This corresponds to taking a series of rational approximations to $\alpha$ which converge $p/q \to \alpha$ at large $q$. Away from criticality, the finite-$q$ approximation $\Gamma(q,\delta)$ converges exponentially $\log |\Gamma(q,\delta) - \Gamma(\infty,\delta)| = O( -q |\delta|^\nu)$~\cite{szabo2018non}. For almost all $\alpha$ in the vicinity of the critical point, $\Gamma$ scales
\begin{equation}
    \Gamma(q,\delta) \sim q^{2-z(\zeta,q)}\mathcal{G}(q^{1/\nu}\delta).
    \label{eq:gamma_scaling}
\end{equation}
Eq.~\eqref{eq:gamma_scaling0} follows from~\eqref{eq:gamma_scaling} by taking the limit of $q \to \infty$ using that the critical scaling ceases for $q \gtrsim \xi$.

We now explain how~\eqref{eq:gamma_scaling} is obtained. At the critical point, the energy of the lowest band varies by $\Delta E = W_0$ across a range of momentum $\Delta k_x = 1/q$, yielding a curvature of $\Gamma(q,0) = O(W_0 q^2)$. We calculate the scaling of $W_0 q^2$ using the RG rules~\eqref{eq:RG} and projecting into the lower Hofstadter band at every RG step.

\begin{figure}[t!]
    \centering
    \includegraphics[width=0.95\linewidth]{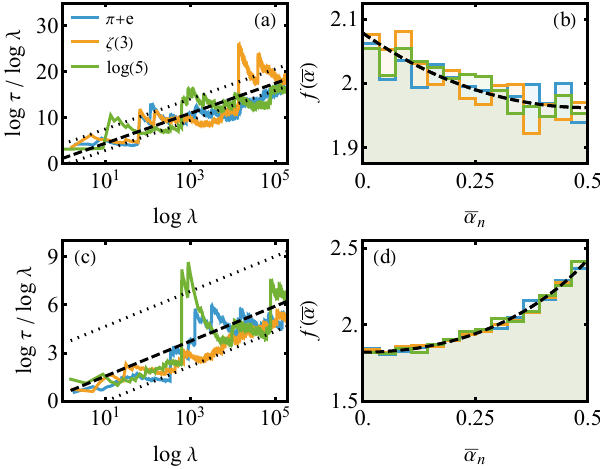}
    \caption{
    \emph{Non-power-law scaling}: The time and length rescalings $\tau$ and $\lambda$ are calculated from the RG~\eqref{eq:RG} after each RG step for $n=10^5$ steps (treating~\eqref{eq:V_scaling} as an equality). (a,b) shows the lower band RG, and (c,d) the central band RG, for 3 irrational $\alpha$ (legend inset). Plot (a) shows agreement with the predicted scaling $\log \tau = \zeta \log \lambda \log \log \lambda$ with $\zeta = 48 G/\pi^3$ (black, dashed). Fluctuations should be compared with the asymptotic 50\% confidence intervals (dotted). (b) shows the convergence of $f_n(\talpha)$ to~\eqref{eq:f}. (c,d) correspond to the central band RG with $\zeta = 16 G / \pi^3$ and $f_n(\talpha)$ converging to~\eqref{eq:f2}.
    }
    \label{Fig:RG}
\end{figure}

We obtain exact results for the RG by using that the flux renormalization map, $B(\talpha)$, is ergodic. Let $\talpha_n = B^n(\talpha)$ be the sequence of renormalized flux densities. Ergodicity entails the $\talpha_n$ approach the same distribution for almost all $\talpha$ (see Fig.~\ref{Fig:RG}b)
\begin{equation}
    f_n(\talpha) := \frac{1}{n}\sum_{m = 0}^{n-1} \delta(\talpha-\talpha_m) \sim f(\talpha) = \frac{\varphi^3/\log \varphi}{\varphi^3 + \talpha-\talpha^2}
    \label{eq:f}
\end{equation}
where $\sim$ indicates convergence in distribution with $n$, and $\varphi = (1 + \sqrt{5})/2$ is the golden ratio (see App.~\ref{app:T}).

The ergodicity of $B$ leads to identical dynamical scaling for almost all $\talpha$. Consider the length rescaling after $n$ RG steps: $\lambda : = a_n/a_0 = \lambda_1 \lambda_2 \cdots$, with $\lambda_m = \lambda(\talpha_m)$
\begin{equation}
    \log \lambda = \sum_{m=0}^{n-1} \log \lambda_m \sim  n \int_0^{1/2}\!\!\!\! d \talpha f(\talpha) \log \lambda( \talpha) = \frac{\pi^2 n }{12 \log \varphi}.
    \label{eq:a_solution}
\end{equation}
We similarly obtain an asymptotically exact time rescaling $\tau := V_0/V_n = \tau_1 \tau_2 \cdots$ with $\tau_n = V_{n-1}/V_n$
\begin{equation}
    \log \tau \sim \frac{4 G}{\pi} \sum_{m=0}^{n-1} s(\talpha_m) = \frac{4 G n}{\pi} \int_0^{1/2} d\talpha f_n (\talpha) s(\talpha) .
    \label{eq:V_solution}
\end{equation}
The asymptotic equality (denoted $\sim$) follows as the sum~\eqref{eq:V_solution} is dominated by small $\talpha_m$ where the RG step is asymptotically exact. However, the integral in~\eqref{eq:V_solution} does not converge as $n\to\infty$. Rather, as $f(\talpha)$ is finite at $\talpha=0$, and $\min_{m<n}\talpha_m = \Theta(n^{-1})$ in measure, the integral diverges as $\sim f(0)\log n$. Specifically,~\cite{aaronson2001local,heinrich1987rates,gouezel2010characterization}
\begin{equation}
    \log \tau = \frac{4 G \, n}{\pi \log \varphi}\left( \log n + \xi_n \right)
    \label{eq:V_solutionb}
\end{equation}
where $\xi_n$ is a sub-leading, heavy tailed (Landau distributed) fluctuation term (see App.~\ref{app:flucs}).

The dynamical scaling,~\eqref{eq:dyn_scaling} with $\zeta = 48 G / \pi^3$, is then obtained from~\eqref{eq:a_solution} and~\eqref{eq:V_solutionb}. The scaling~\eqref{eq:gamma_scaling} at $\delta = 0$ follows from $\Gamma(q,0) = O(W_0 q^2)$ with $q = O(a_n)$ and $W_0 = O(V_n)$. The $\delta$-dependence is fixed by~\eqref{eq:xi}. 

Fig.~\ref{Fig:RG}a shows the scaling of $\Gamma$ for several $\alpha$, agreeing with the universal form~\eqref{eq:dyn_scaling} (dashed). Visible deviations are consistent with the sub-leading but heavy tailed fluctuations predicted~\eqref{eq:dyn_scaling} (see App.~\ref{app:flucs}). 

The non-power-law scaling obtained here originates from rare RG steps in which the renormalized flux is small (i.e. $H$ is close to commensurate) and hence the dynamical exponent $z_n$ is large. Specifically, $z_n = \log \tau_n / \log \lambda_n \sim 4 G/|\pi\talpha_n \log \talpha_n|$ diverges at small $\talpha_n$. 

\paragraph*{\textbf{Specific heat of Fermi gas at half-filling:}} The non-power-law scaling is not unique to the lowest band RG. For example, consider promoting~\eqref{eq:AA} to a system of non-interacting fermions at half-filling. This system has specific heat
\begin{equation}
    c = \frac{d u}{d T} = \frac{d}{d T} \int_{-\infty}^{\infty} \!\!\! d E \, E \, \rho(E) \, n_\mathrm{F}(E)
    \label{eq:CV}
\end{equation}
where $u$ is the energy density, $n_\mathrm{F}(E) = 1/(\e^{E/T}+1)$ is the Fermi-Dirac distribution, and $\rho(E)$ is the density of states (DOS) of $H_\mathrm{AA}$. The low-temperature $T$ behaviour of $c$ is set by the scaling of $\rho(E)$ as $E \to 0$. At the critical point the integrated DOS, $N(E) = \int_{-E}^{E} \rho(E') d E'$, has scaling behaviour 
\begin{equation}
    N(\tau E) \sim \tau^{1/z(\zeta,\tau)} N(E)
    \label{eq:rho_lowT}
\end{equation}
where $\zeta = 16 G / \pi^3$, and $\sim$ indicates asymptotic equality at small $\tau$. Substituting $E = x T$ into~\eqref{eq:CV} and differentiating~\eqref{eq:rho_lowT} to write $\rho(x T) \sim T^{-1 + 1/z(\zeta,\tau)} \rho(x)$, we scale out the $T$-dependence to obtain the low-$T$ specific heat scaling
\begin{equation}
    c \sim k \, T^{1/z(\zeta,T)} , \qquad \zeta = 16 G /\pi^3.
    \label{eq:C_lowT}
\end{equation}
for $T$-independent $k$.

We obtain the low-energy DOS scaling~\eqref{eq:rho_lowT} from the RG~\eqref{eq:RG}, projecting into the central band at each step. After $n$ steps, we have a renormalized energy and length scales $V_n$, and $a_n$ which determine the low-energy behaviour of the DOS as $N(4 V_n) = a_0/a_n$.

The RG for the central band proceeds according to~\eqref{eq:RG} with $s(z) = z\lambda(z) = z/(1-2z)$. To obtain analytic results, instead of taking single RG steps, we take ``RG super-steps'' consisting of $r$ RG steps, with $r(\talpha) := \lfloor 1/(2 \talpha) \rfloor$ determined by $\talpha$~\footnote{RG super-steps address the pathology that most RG steps result in negligible rescaling of length and time.}. Under one super-step the renormalization rules~\eqref{eq:RG} hold with $\lambda(z) = (1 - 2 z r(z))^{-1}$ and $s(z) = \sum_{p=1}^r(1-2pz)^{-1}$. The analysis then proceeds as in the previous section. The super-step flux renormalization map $B(\talpha)$ is ergodic, thus the $\talpha_m : = B^m(\talpha)$ converge to a steady state distribution
\begin{equation}
    f_n(\talpha) := \frac{1}{n}\sum_{m = 0}^{n-1} \delta(\talpha-\talpha_m) \sim f(\talpha) = \frac{2}{\log 3}\,\frac{1}{1 - \talpha^2}.
    \label{eq:f2}
\end{equation}
For a central band RG super-step, the dynamical exponent $z_n = z(\talpha_n)$ diverges as $\talpha_n \to 1/(2p)$ for integer $p$ (rather than simply as $\talpha \to 0$ as before). Otherwise, the analysis proceeds by direct generalization, yielding length and time rescalings
\begin{align}
    \log \lambda = \log \frac{a_n}{a_0} \sim \frac{\pi^2 n}{4 \log 3}, \quad \log \tau = \log \frac{V_0}{V_n} \sim \frac{4 G n \log n}{\pi \log 3} \label{eq:aV_solution2}
\end{align}
from which we obtain the dynamical scaling~\eqref{eq:dyn_scaling} (see Fig.~\ref{Fig:RG}c). Combining~\eqref{eq:aV_solution2} and $N(4 V_n) = a_0/a_n$ we obtain the low-energy scaling of the DOS~\eqref{eq:rho_lowT} and hence the specific heat scaling~\eqref{eq:C_lowT}.

\paragraph*{\textbf{Discussion:}} We have shown, in the critical HH model, for almost all $\alpha$, the scaling of the lowest bandwidth~\eqref{eq:Gam}, and the heat capacity of a half-filled Fermi gas~\eqref{eq:CV} both exhibit the non-power-law time-length scaling~\eqref{eq:dyn_scaling} with respective coefficients $\zeta = 48G/\pi^3$, and $\zeta = 16 G / \pi^3$ where $G$ is Catalan's constant. 

Our results apply for almost all $\alpha$. This extends previous work~\cite{han1994critical,kohmoto1986quasiperiodic,fujiwara1989multifractal,hiramoto1988dynamics,hiramoto1988dynamicsII,hiramoto1992electronic,kohmoto1987critical,hiramoto1992electronic,gopalakrishnan2017self,devakul2017anderson,levitov1989renormalization,kohmoto1983localization,ostlund1984renormalization,wurtz1988renormalization,ceccatto1989quasiperiodic,benza1989quantum,luck1993critical,chandran2017localization,crowley2018quasi,crowley2018critical,agrawal2020universality}, which considered specific $\alpha$, often quadratic integers (e.g. metallic ratios). For quadratic integer $\alpha$, the orbit $\talpha_n$ is periodic. Such $\alpha$ are instances of the measure zero exceptions to our result~\eqref{eq:dyn_scaling}, and yield power-law dynamical scaling.

The RG employed is generically approximate for a single step. However, it is asymptotically exact in the respective limits $\talpha_n \to 0, \tfrac12$ for the two cases studied. As these limits dominate the dynamical scaling, our result is asymptotically exact. The deviations apparent in Fig.~\ref{Fig:RG} correspond to the subleading corrections in~\eqref{eq:dyn_scaling}.

Two key features of the RG underpin the $\alpha$-independent non-power-law scaling~\eqref{eq:dyn_scaling}: (i) $\talpha$ is renormalized by an ergodic map $B$, causing the distribution of $\talpha_n$ to converge to an $\alpha$-independent distribution $f(\talpha)$ and (ii) the single step dynamical exponent $z_n = z(\talpha_n)$ where $z(\alpha)$ has a non-integrable divergence, causing rare large $z_n$ to dominate the dynamical scaling.

We expect the scaling~\eqref{eq:dyn_scaling} to be generic for a broad class of critical models, though with model-dependent $\zeta$. Specifically, consider a generalized-HH model $H_\mathrm{G}=V_s(\hat{x},\hat{y})$ where $V_s$ is any smooth potential periodic in its two arguments, and $s$ is a tuning parameter. Generically, a finite fraction of the equipotentials of $V_s$ are open, and all open equipotentials extend parallel to a single lattice vector $\vec{b}$. For critical values of $s$, $\vec{b}$ changes discretely, and $H_\mathrm{G}$ generalizes the critical HH model. The spectrum of the critical generalized-HH model is also fractal, and an analogous RG transformation may be constructed. Under this RG we expect dynamical scaling~\eqref{eq:dyn_scaling} with an $\alpha$-independent value of $\zeta$ provided it preserves the properties (i) and (ii) above. Property (i) is anticipated as the RG transformation necessarily geometrically amplifies small changes to the initial value of $\alpha$, i.e. it is always chaotic, and we expect also ergodic. Property (ii) follows by direct generalization of the WKB methods used in the HH model~\cite{szabo2018non,wilkinson1984critical,wilkinson1987exact,han1994critical} (see App.~\ref{app:WKB}).

Lastly, as the $\alpha$-independence of the scaling stems from Gauss-Kuzmin statistics (with $B(\talpha)$ replacing the usual Gauss map), and not model-specific features, we conjecture that this is generic of incommensurate systems.

\paragraph*{\textbf{Acknowledgments}:} We are grateful for useful discussions with A. Szab\'{o}, and to C. Murthy for useful comments on a draft. P.C. was supported by the NSF STC ``Center for Integrated Quantum Materials'' under Cooperative Agreement No. DMR-1231319. 

\bibliographystyle{apsrev4-2}
\bibliography{bib}

@article{wannier1978result,
  title={A result not dependent on rationality for Bloch electrons in a magnetic field},
  author={Wannier, GH},
  journal={physica status solidi (b)},
  volume={88},
  number={2},
  pages={757--765},
  year={1978},
  publisher={Wiley Online Library}
}

@article{hofstadter1976energy,
  title={Energy levels and wave functions of {Bloch} electrons in rational and irrational magnetic fields},
  author={Hofstadter, Douglas R},
  journal={Physical review B},
  volume={14},
  number={6},
  pages={2239},
  year={1976},
  publisher={APS}
}

@article{rudinger1997hofstadter,
  title={Hofstadter rules and generalized dimensions of the spectrum of Harper's equation},
  author={R{\"u}dinger, Andreas and Pi{\'e}chon, Fr{\'e}d{\'e}ric},
  journal={Journal of Physics A: Mathematical and General},
  volume={30},
  number={1},
  pages={117},
  year={1997},
  publisher={IOP Publishing}
}

@article{szabo2018non,
  title={Non-power-law universality in one-dimensional quasicrystals},
  author={Szab{\'o}, Attila and Schneider, Ulrich},
  journal={Physical Review B},
  volume={98},
  number={13},
  pages={134201},
  year={2018},
  publisher={APS}
}

@article{simon1982almost,
  title={Almost periodic Schr{\"o}dinger operators: a review},
  author={Simon, Barry},
  journal={Advances in Applied Mathematics},
  volume={3},
  number={4},
  pages={463--490},
  year={1982},
  publisher={Elsevier}
}

@article{thiem2015origin,
  title={Origin of the log-periodic oscillations in the quantum dynamics of electrons in quasiperiodic systems},
  author={Thiem, Stefanie},
  journal={Philosophical Magazine},
  volume={95},
  number={11},
  pages={1233--1243},
  year={2015},
  publisher={Taylor \& Francis}
}

@article{damanik2009spectrum,
  title={The spectrum of the almost Mathieu operator},
  author={Damanik, David},
  journal={arXiv preprint arXiv:0908.1093},
  year={2009}
}

@article{aubry1980analyticity,
	Author = {Aubry, S. and Andr\'e, G.},
	Journal = {Ann. Israel Phys. Soc},
	Pages = {133},
	Title = {Analyticity breaking and Anderson localization in incommensurate lattices},
	Volume = {3},
	Year = {1980}}

@article{azbel1979quantum,
	Author = {Azbel, M. Ya.},
	Issue = {26},
	Journal = {Phys. Rev. Lett.},
	Month = {Dec},
	Numpages = {0},
	Pages = {1954--1957},
	Publisher = {American Physical Society},
	Title = {Quantum Particle in One-Dimensional Potentials with Incommensurate Periods},
	Volume = {43},
	Year = {1979},
}

@article{thouless1990scaling,
  title={Scaling for the discrete Mathieu equation},
  author={Thouless, DJ},
  journal={Communications in mathematical physics},
  volume={127},
  number={1},
  pages={187--193},
  year={1990},
  publisher={Springer}
}

@article{ketzmerick1997determines,
  title={What determines the spreading of a wave packet?},
  author={Ketzmerick, R and Kruse, K and Kraut, S and Geisel, T},
  journal={Physical review letters},
  volume={79},
  number={11},
  pages={1959},
  year={1997},
  publisher={APS}
}

@article{ostlund1984renormalization,
  title={Renormalization-group analysis of the discrete quasiperiodic Schr{\"o}dinger equation},
  author={Ostlund, Stellan and Pandit, Rahul},
  journal={Physical Review B},
  volume={29},
  number={3},
  pages={1394},
  year={1984},
  publisher={APS}
}

@article{hiramoto1988dynamicsII,
  title={Dynamics of an Electron in Quasiperiodic Systems. II. Harper's Model},
  author={Hiramoto, Hisashi and Abe, Shuji},
  journal={Journal of the Physical Society of Japan},
  volume={57},
  number={4},
  pages={1365--1371},
  year={1988},
  publisher={The Physical Society of Japan}
}

@article{hiramoto1988dynamics,
  title={Dynamics of an electron in quasiperiodic systems. I. Fibonacci model},
  author={Hiramoto, Hisashi and Abe, Shuji},
  journal={Journal of the Physical Society of Japan},
  volume={57},
  number={1},
  pages={230--240},
  year={1988},
  publisher={The Physical Society of Japan}
}

@article{fujiwara1989multifractal,
  title={Multifractal wave functions on a Fibonacci lattice},
  author={Fujiwara, T and Kohmoto, Mahito and Tokihiro, T},
  journal={Physical Review B},
  volume={40},
  number={10},
  pages={7413},
  year={1989},
  publisher={APS}
}

@article{kohmoto1986quasiperiodic,
  title={Quasiperiodic lattice: Electronic properties, phonon properties, and diffusion},
  author={Kohmoto, Mahito and Banavar, Jayanth R},
  journal={Physical Review B},
  volume={34},
  number={2},
  pages={563},
  year={1986},
  publisher={APS}
}

@article{chandran2017localization,
	Author = {Chandran, A. and Laumann, C. R.},
	Issue = {3},
	Journal = {Phys. Rev. X},
	Month = {Sep},
	Numpages = {21},
	Pages = {031061},
	Publisher = {American Physical Society},
	Title = {Localization and Symmetry Breaking in the Quantum Quasiperiodic Ising Glass},
	Volume = {7},
	Year = {2017}
}

@article{luck1993critical,
	Author = {Luck, J.M.},
	Issn = {0022-4715},
	Journal = {Journal of Statistical Physics},
	Language = {English},
	Number = {3-4},
	Pages = {417-458},
	Publisher = {Kluwer Academic Publishers-Plenum Publishers},
	Title = {Critical behavior of the aperiodic quantum Ising chain in a transverse magnetic field},
	Volume = {72},
	Year = {1993}
}

@book{khinchin1963continued,
  title={Continued Fractions},
  author={Khinchin, A. Ya.},
  year={1963},
  publisher={P. Noordhoff},
  address={Groningen},
  note={Translated by P. Wynn}
}

@article{hermisson1997aperiodic,
  title={Aperiodic Ising quantum chains},
  author={Hermisson, Joachim and Grimm, Uwe and Baake, Michael},
  journal={Journal of Physics A: Mathematical and General},
  volume={30},
  number={21},
  pages={7315},
  year={1997},
  publisher={IOP Publishing}
}

@article{vieira2005aperiodic,
  title={Aperiodic quantum XXZ chains: Renormalization-group results},
  author={Vieira, Andr{\'e} P},
  journal={Physical Review B},
  volume={71},
  number={13},
  pages={134408},
  year={2005},
  publisher={APS}
}

@article{han1994critical,
  title={Critical and bicritical properties of Harper’s equation with next-nearest-neighbor coupling},
  author={Han, JH and Thouless, DJ and Hiramoto, H and Kohmoto, M},
  journal={Physical Review B},
  volume={50},
  number={16},
  pages={11365},
  year={1994},
  publisher={APS}
}

@article{wilkinson1984critical,
  title={Critical properties of electron eigenstates in incommensurate systems},
  author={Wilkinson, M},
  journal={Proceedings of the Royal Society of London. A. Mathematical and Physical Sciences},
  volume={391},
  number={1801},
  pages={305--350},
  year={1984},
  publisher={The Royal Society London}
}

@article{thouless1983bandwidths,
  title={Bandwidths for a quasiperiodic tight-binding model},
  author={Thouless, DJ},
  journal={Physical Review B},
  volume={28},
  number={8},
  pages={4272},
  year={1983},
  publisher={APS}
}

@article{hiramoto1992electronic,
  title={Electronic spectral and wavefunction properties of one-dimensional quasiperiodic systems: a scaling approach},
  author={Hiramoto, Hisashi and Kohmoto, Mahito},
  journal={International Journal of Modern Physics B},
  volume={06},
  number={03n04},
  pages={281--320},
  year={1992},
  publisher={World Scientific}
}

@article{kohmoto1987critical,
  title={Critical wave functions and a Cantor-set spectrum of a one-dimensional quasicrystal model},
  author={Kohmoto, Mahito and Sutherland, Bill and Tang, Chao},
  journal={Physical Review B},
  volume={35},
  number={3},
  pages={1020},
  year={1987},
  publisher={APS}
}

@incollection{avila2006solving,
  title={Solving the ten martini problem},
  author={Avila, Artur and Jitomirskaya, Svetlana},
  booktitle={Mathematical physics of quantum mechanics},
  pages={5--16},
  year={2006},
  publisher={Springer}
}

@article{wurtz1988renormalization,
  title={Renormalization-group study of Fibonacci chains},
  author={W{\"u}rtz, D and Schneider, T and Politi, A},
  journal={Physics Letters A},
  volume={129},
  number={2},
  pages={88--92},
  year={1988},
  publisher={Elsevier}
}

@article{gopalakrishnan2017self,
  title={Self-dual quasiperiodic systems with power-law hopping},
  author={Gopalakrishnan, Sarang},
  journal={Physical Review B},
  volume={96},
  number={5},
  pages={054202},
  year={2017},
  publisher={APS}
}

@article{ceccatto1989quasiperiodic,
  title={Quasiperiodic Ising model in a transverse field: Analytical results},
  author={Ceccatto, HA},
  journal={Physical review letters},
  volume={62},
  number={2},
  pages={203},
  year={1989},
  publisher={APS}
}

@article{benza1989quantum,
  title={Quantum Ising quasi-crystal},
  author={Benza, VG},
  journal={EPL (Europhysics Letters)},
  volume={8},
  number={4},
  pages={321},
  year={1989},
  publisher={IOP Publishing}
}

@article{devakul2017anderson,
  title={Anderson localization transitions with and without random potentials},
  author={Devakul, Trithep and Huse, David A},
  journal={Physical Review B},
  volume={96},
  number={21},
  pages={214201},
  year={2017},
  publisher={APS}
}

@article{crowley2018quasi,
  title={Quasiperiodic Quantum Ising Transitions in 1D},
  author={Crowley, PJD and Chandran, A and Laumann, CR},
  journal={Physical review letters},
  volume={120},
  number={17},
  pages={175702},
  year={2018},
  publisher={APS}
}

@article{suslov1982localization,
  title={Localization in one-dimensional incommensurate systems},
  author={Suslov, IM},
  journal={Zh. Eksp. Teor. Fiz},
  volume={83},
  pages={1079--1088},
  year={1982}
}

@article{hunt2013massive,
  title={Massive Dirac fermions and Hofstadter butterfly in a van der Waals heterostructure},
  author={Hunt, Benjamin and Sanchez-Yamagishi, Javier D and Young, Andrea F and Yankowitz, Matthew and LeRoy, Brian J and Watanabe, Kenji and Taniguchi, Takashi and Moon, Pilkyung and Koshino, Mikito and Jarillo-Herrero, Pablo and others},
  journal={Science},
  volume={340},
  number={6139},
  pages={1427--1430},
  year={2013},
  publisher={American Association for the Advancement of Science}
}

@article{dean2013hofstadter,
  title={Hofstadter’s butterfly and the fractal quantum Hall effect in moir{\'e} superlattices},
  author={Dean, Cory R and Wang, L and Maher, P and Forsythe, C and Ghahari, Fereshte and Gao, Y and Katoch, Jyoti and Ishigami, M and Moon, P and Koshino, M and others},
  journal={Nature},
  volume={497},
  number={7451},
  pages={598--602},
  year={2013},
  publisher={Nature Publishing Group}
}

@article{carr2017twistronics,
  title={Twistronics: Manipulating the electronic properties of two-dimensional layered structures through their twist angle},
  author={Carr, Stephen and Massatt, Daniel and Fang, Shiang and Cazeaux, Paul and Luskin, Mitchell and Kaxiras, Efthimios},
  journal={Physical Review B},
  volume={95},
  number={7},
  pages={075420},
  year={2017},
  publisher={APS}
}

@article{andrei2020graphene,
  title={Graphene bilayers with a twist},
  author={Andrei, Eva Y and MacDonald, Allan H},
  journal={Nature materials},
  volume={19},
  number={12},
  pages={1265--1275},
  year={2020},
  publisher={Nature Publishing Group}
}

@article{bordia2017probing,
  title={Probing slow relaxation and many-body localization in two-dimensional quasiperiodic systems},
  author={Bordia, Pranjal and L{\"u}schen, Henrik and Scherg, Sebastian and Gopalakrishnan, Sarang and Knap, Michael and Schneider, Ulrich and Bloch, Immanuel},
  journal={Physical Review X},
  volume={7},
  number={4},
  pages={041047},
  year={2017},
  publisher={APS}
}

@article{schreiber2015observation,
  title={Observation of many-body localization of interacting fermions in a quasirandom optical lattice},
  author={Schreiber, Michael and Hodgman, Sean S and Bordia, Pranjal and L{\"u}schen, Henrik P and Fischer, Mark H and Vosk, Ronen and Altman, Ehud and Schneider, Ulrich and Bloch, Immanuel},
  journal={Science},
  volume={349},
  number={6250},
  pages={842--845},
  year={2015},
  publisher={American Association for the Advancement of Science}
}

@book{satija2016butterfly,
  title={The Butterfly in the Quantum World: The story of the most fascinating quantum fractal},
  author={Satija, Indubala I},
  year={2016},
  publisher={Morgan \& Claypool Publishers},
  address={San Rafael, CA}
}

@article{azbel1964zh,
  title={Energy spectrum of a conduction electron in a magnetic field},
  author={Azbel, M. Ya.},
  journal={Sov. Phys. JETP},
  volume={19},
  pages={634},
  year={1964},
  note={[Zh. Eksp. Teor. Fiz. \textbf{46}, 929 (1964)]}
}

@article{agrawal2020universality,
  title={Universality and quantum criticality in quasiperiodic spin chains},
  author={Agrawal, Utkarsh and Gopalakrishnan, Sarang and Vasseur, Romain},
  journal={Nature Communications},
  volume={11},
  pages={2225},
  year={2020},
  doi={10.1038/s41467-020-15760-5}
}

@article{levitov1989renormalization,
  title={Renormalization group for a quasiperiodic Schr{\"o}dinger operator},
  author={Levitov, L. S.},
  journal={J. Phys. (France)},
  volume={50},
  number={7},
  pages={707--716},
  year={1989},
  doi={10.1051/jphys:01989005007070700}
}

@article{kohmoto1983localization,
  title={Localization problem in one dimension: Mapping and escape},
  author={Kohmoto, Mahito and Kadanoff, Leo P and Tang, Chao},
  journal={Physical Review Letters},
  volume={50},
  number={23},
  pages={1870},
  year={1983},
  publisher={APS}
}

@book{goldenfeld2018lectures,
  title={Lectures on phase transitions and the renormalization group},
  author={Goldenfeld, Nigel},
  year={2018},
  publisher={CRC Press}
}

@incollection{lieb2002superfluidity,
  title={Superfluidity in dilute trapped Bose gases},
  author={Lieb, Elliott H and Seiringer, Robert and Yngvason, Jakob},
  booktitle={The Stability of Matter: From Atoms to Stars},
  pages={903--908},
  year={2002},
  publisher={Springer}
}

@article{roth2003phase,
  title={Phase diagram of bosonic atoms in two-color superlattices},
  author={Roth, Robert and Burnett, Keith},
  journal={Physical Review A},
  volume={68},
  number={2},
  pages={023604},
  year={2003},
  publisher={APS}
}

@article{cestari2010finite,
  title={Finite-size effects in Anderson localization of one-dimensional Bose-Einstein condensates},
  author={Cestari, Jardel Caminha Carvalho and Foerster, Angela and Gusmao, MA},
  journal={Physical Review A},
  volume={82},
  number={6},
  pages={063634},
  year={2010},
  publisher={APS}
}

@article{bellissard1982cantor,
  title={Cantor spectrum for the almost Mathieu equation},
  author={Bellissard, Jean and Simon, Barry},
  journal={Journal of functional analysis},
  volume={48},
  number={3},
  pages={408--419},
  year={1982},
  publisher={Elsevier}
}

@article{chambers1965linear,
  title={Linear-network model for magnetic breakdown in two dimensions},
  author={Chambers, WG},
  journal={Physical Review},
  volume={140},
  number={1A},
  pages={A135},
  year={1965},
  publisher={APS}
}

@article{last1994zero,
  title={Zero measure spectrum for the almost Mathieu operator},
  author={Last, Yoram},
  journal={Communications in mathematical physics},
  volume={164},
  number={2},
  pages={421--432},
  year={1994},
  publisher={Springer}
}

@article{helffersjostrand1988,
  author = {Helffer, B. and Sj\"ostrand, J.},
  title = {Analyse semi-classique pour l'\'equation de {Harper} (avec application \`a l'\'equation de {Schr\"odinger} avec champ magn\'etique)},
  journal = {M\'em. Soc. Math. Fr.},
  volume = {34},
  pages = {1--113},
  year = {1988},
  doi = {10.24033/msmf.335}
}

@article{helffersjostrand1989,
  author = {Helffer, B. and Sj\"ostrand, J.},
  title = {Semi-classical analysis for {Harper's} equation. {III}: {Cantor} structure of the spectrum},
  journal = {M\'em. Soc. Math. Fr.},
  volume = {39},
  pages = {1--124},
  year = {1989},
  doi = {10.24033/msmf.346}
}

@article{helffersjostrand1990,
  author = {Helffer, B. and Sj\"ostrand, J.},
  title = {Analyse semi-classique pour l'\'equation de {Harper.} {II}: comportement semi-classique pr\`es d'un rationnel},
  journal = {M\'em. Soc. Math. Fr.},
  volume = {40},
  pages = {1--139},
  year = {1990},
  doi = {10.24033/msmf.347}
}

@article{aaronson2001local,
  title={Local limit theorems for partial sums of stationary sequences generated by Gibbs--Markov maps},
  author={Aaronson, Jon and Denker, Manfred},
  journal={Stochastics and Dynamics},
  volume={1},
  number={02},
  pages={193--237},
  year={2001},
  publisher={World Scientific}
}

@article{gouezel2010characterization,
  title={Characterization of weak convergence of Birkhoff sums for Gibbs-Markov maps},
  author={Gou{\"e}zel, S{\'e}bastien},
  journal={Israel Journal of Mathematics},
  volume={180},
  number={1},
  pages={1--41},
  year={2010},
  publisher={Springer}
}

@article{heinrich1987rates,
  title={Rates of Convergence in Stable Limit Theorems for Sums of Exponentially $\Psi$-mixing Random Variables with an Application to Metric Theory of Continued Fractions},
  author={Heinrich, Lothar},
  journal={Mathematische Nachrichten},
  volume={131},
  number={1},
  pages={149--165},
  year={1987},
  publisher={Wiley Online Library}
}

@article{tchoumakov2016magnetic,
  title={Magnetic-field-induced relativistic properties in type-I and type-II Weyl semimetals},
  author={Tchoumakov, Serguei and Civelli, Marcello and Goerbig, Mark O},
  journal={Physical review letters},
  volume={117},
  number={8},
  pages={086402},
  year={2016},
  publisher={APS}
}

@article{berry1972semiclassical,
  title={Semiclassical approximations in wave mechanics},
  author={Berry, Michael V and Mount, KE},
  journal={Reports on Progress in Physics},
  volume={35},
  number={1},
  pages={315--397},
  year={1972}
}

@inbook{bellissard1989cstar, place={Cambridge}, series={London Mathematical Society Lecture Note Series}, title={C*-algebras in solid state physics: 2D electrons in uniform magnetic field}, booktitle={Operator Algebras and Applications}, publisher={Cambridge University Press}, author={Bellissard, J.}, editor={Evans, David E. and Takesaki, MasamichiEditors}, year={1989}, pages={49–76}, collection={London Mathematical Society Lecture Note Series}}

@article{jitomirskaya1999metal,
  title={Metal-insulator transition for the almost {Mathieu} operator},
  author={Jitomirskaya, Svetlana Ya.},
  journal={Annals of Mathematics},
  volume={150},
  number={3},
  pages={1159--1175},
  year={1999},
  doi={10.2307/121066}
}

@book{landau1981quantum,
  author = {Landau, L. D. and Lifshitz, E. M.},
  title = {Quantum Mechanics: Non-Relativistic Theory},
  series = {Course of Theoretical Physics},
  volume = {3},
  edition = {3},
  publisher = {Pergamon},
  address = {Oxford},
  year = {1977}
}

@article{uri2023superconductivity,
  title={Superconductivity and strong interactions in a tunable moir$\backslash$'e quasiperiodic crystal},
  author={Uri, Aviram and de la Barrera, Sergio C and Randeria, Mallika T and Rodan-Legrain, Daniel and Devakul, Trithep and Crowley, Philip JD and Paul, Nisarga and Watanabe, Kenji and Taniguchi, Takashi and Lifshitz, Ron and others},
  journal={arXiv preprint arXiv:2302.00686},
  year={2023}
}

@article{crowley2018critical,
  title={Critical behaviour of the quasi-periodic quantum Ising chain},
  author={Crowley, PJD and Laumann, CR and Chandran, A},
  journal={Journal of Statistical Mechanics: Theory and Experiment},
  volume={2022},
  number={8},
  pages={083102},
  year={2022},
  publisher={IOP Publishing}
}

@article{wilkinson1987exact,
  title={An exact renormalisation group for Bloch electrons in a magnetic field},
  author={Wilkinson, Michael},
  journal={Journal of Physics A: Mathematical and General},
  volume={20},
  number={13},
  pages={4337},
  year={1987},
  publisher={IOP Publishing}
}

@article{thouless1983wavefunction,
  title={Wavefunction scaling in a quasi-periodic potential},
  author={Thouless, DJ and Niu, Q},
  journal={Journal of Physics A: Mathematical and General},
  volume={16},
  number={9},
  pages={1911},
  year={1983},
  publisher={IOP Publishing}
}

@article{zilberman1956zh,
  author={Zilberman, G. E.},
  journal={Sov. Phys. JETP},
  volume={3},
  pages={835},
  year={1956},
  note={[Zh. Eksp. Teor. Fiz. \textbf{30}, 1092 (1956)]}
}

@article{rauh1974bloch,
  title={Bloch electrons in irrational magnetic fields},
  author={Rauh, A and Wannier, GH and Obermair, G},
  journal={physica status solidi (b)},
  volume={63},
  number={1},
  pages={215--229},
  year={1974},
  publisher={Wiley Online Library}
}

@article{paul2022moir,
  title={Moir$\backslash$'e Landau fans and magic zeros},
  author={Paul, Nisarga and Crowley, Philip JD and Devakul, Trithep and Fu, Liang},
  journal={arXiv preprint arXiv:2202.05854},
  year={2022}
}

@article{briggs2003precise,
  title={A precise computation of the Gauss--Kuzmin--Wirsing constant},
  author={Briggs, Keith},
  year={2003},
  journal={unpublished},
  note={preprint available at \url{http://keithbriggs.info/documents/wirsing.pdf}}
}

@article{flajolet1995gauss,
  title={On the Gauss-Kuzmin-Wirsing Constant},
  author={Flajolet, P. and Vallée, B},
  year={1995},
  journal={unpublished},
  note={preprint available at \url{http://algo.inria.fr/flajolet/Publications/gauss-kuzmin.ps}}
}

@article{schultka1994finite,
  title={Finite-size scaling in two-dimensional superfluids},
  author={Schultka, Norbert and Manousakis, Efstratios},
  journal={Physical Review B},
  volume={49},
  number={17},
  pages={12071},
  year={1994},
  publisher={APS}
}

@book{cardy1996scaling,
  title={Scaling and renormalization in statistical physics},
  author={Cardy, John},
  volume={5},
  year={1996},
  publisher={Cambridge university press}
}

@article{ponomarenko2013cloning,
  title={Cloning of Dirac fermions in graphene superlattices},
  author={Ponomarenko, LA and Gorbachev, RV and Yu, GL and Elias, DC and Jalil, R and Patel, AA and Mishchenko, A and Mayorov, AS and Woods, CR and Wallbank, JR and others},
  journal={Nature},
  volume={497},
  number={7451},
  pages={594--597},
  year={2013},
  publisher={Nature Publishing Group}
}

@article{deissler2010delocalization,
  title={Delocalization of a disordered bosonic system by repulsive interactions},
  author={Deissler, Benjamin and Zaccanti, Matteo and Roati, Giacomo and D’Errico, Chiara and Fattori, Marco and Modugno, Michele and Modugno, Giovanni and Inguscio, Massimo},
  journal={Nature physics},
  volume={6},
  number={5},
  pages={354--358},
  year={2010},
  publisher={Nature Publishing Group}
}

@article{roati2008anderson,
  title={Anderson localization of a non-interacting Bose--Einstein condensate},
  author={Roati, Giacomo and D’Errico, Chiara and Fallani, Leonardo and Fattori, Marco and Fort, Chiara and Zaccanti, Matteo and Modugno, Giovanni and Modugno, Michele and Inguscio, Massimo},
  journal={Nature},
  volume={453},
  number={7197},
  pages={895--898},
  year={2008},
  publisher={Nature Publishing Group}
}

@article{aidelsburger2013realization,
  title={Realization of the Hofstadter Hamiltonian with ultracold atoms in optical lattices},
  author={Aidelsburger, Monika and Atala, Marcos and Lohse, Michael and Barreiro, Julio T and Paredes, B and Bloch, Immanuel},
  journal={Physical review letters},
  volume={111},
  number={18},
  pages={185301},
  year={2013},
  publisher={APS}
}

@article{last1992sum,
  title={A sum rule for the dispersion relations of the rational Harper equation},
  author={Last, Y and Wilkinson, M},
  journal={Journal of Physics A: Mathematical and General},
  volume={25},
  number={22},
  pages={6123},
  year={1992},
  publisher={IOP Publishing}
}

\appendix
\begin{widetext}

\numberwithin{equation}{section}

\begin{center}\textbf{\large Supplemental Material to ``Universal non-power law scaling from a chaotic RG in the Harper-Hofstadter model''}\end{center}

\section{Derivation of Eq.~(6)}
\label{app:chambers}

In this section we derive~\eqref{eq:Ea}, that the energetic dependence of a single band of the HH model for rational flux $\alpha = p/q$ is given by
\begin{equation}
    E_j(x_0,y_0) =  E_{j}^* - 2 V_x' \cos q y_0 - 2 V_y' \cos q x_0 + O(W_j^2/\Delta_j).
\end{equation}
In the related, AA model the quantities $(x_0,y_0)$ have a straightforward interpretation: $x_0$ acts as the phase of the potential, and $y_0$ is a crystal momentum, and \emph{vice versa} in the dual model obtained by writing in the $y$-basis. The results of this section are obtained using well known properties of the characteristic equation and spectrum of the HH model~\cite{chambers1965linear,bellissard1982cantor,thouless1983bandwidths,thouless1990scaling,thouless1990scaling,last1994zero,last1992sum}.

Our first step is to obtain a useful form for the characteristic equation, the roots of which are the bands $E_j(x_0,y_0)$. We begin by noting that, per~\eqref{eq:AA}, when written in the $x$-basis $\hat{x} \ket{x} = x \ket{x}$ the HH hamiltonian takes the form $H  = \int_0^{2\pi\alpha} d x_0 H_\mathrm{AA}(x_0)$ with
\begin{equation}
    \begin{aligned}
    H_\mathrm{AA}(x_0) & =  \sum_{n \in \mathbb{Z}}   \Big[ V_y \left(  \ket{x_0 + 2 \pi \alpha (n + 1) }\bra{x_0 + 2 \pi \alpha n} +\mathrm{h.c} \right) + 2 V_x \cos x_n \ket{x_0 + 2 \pi \alpha n}\bra{x_0 + 2 \pi \alpha n} \Big].
    \end{aligned}
\end{equation}
This Hamiltonian is manifestly periodic under a shift $x \to x + 2 \pi \alpha q$, and so we may project in a momentum sector, in which we obtain the Bloch Hamiltonian
\begin{equation}
    \begin{aligned}
    H_\mathrm{B}(x_0,y_0) & = -  \sum_{n \in \mathbb{Z}}   \Big[ V_y \left( \e^{ i y_0} \ket{n+1 }\bra{n} +\mathrm{h.c} \right) + 2 V_x \cos x_n \ket{n}\bra{n} \Big]
    \end{aligned}
\end{equation}
where we identify $\ket{n} \equiv \ket{n+q}$. The spectrum of $H$ is thus made up of bands $E_j(x_0,y_0)$ determined by the solutions of the characteristic equation 
\begin{equation}
    C(E_j(x_0,y_0),x_0,y_0) = 0
\end{equation}
where $C$ is the characteristic polynomial given by
\begin{equation}
    C(E,x_0,y_0) := \det ( H_\mathrm{B}(x_0,y_0) - E).
\end{equation}
Remarkably, the characteristic polynomial has a very simple dependence on $x_0,y_0$
\begin{equation}
    C(E,x_0,y_0) = P(E) + C_0(x_0,y_0)
\end{equation}
where $P(E)$ is a $q$th order polynomial independent of $x_0$ and $y_0$, and $C_0$ is an energy independent constant
\begin{equation}
    \begin{aligned}
       P(E) & = \det ( H_\mathrm{B}(\pi/2 q,\pi / 2 q) - E)
       \\
       C_0(x_0,y_0) &= - 2 V_x^q \cos q x_0 - 2 V_y^q \cos q y_0.
    \end{aligned}
\end{equation}
Hence the roots of the characteristic polynomial are the solutions to the equation
\begin{equation}
    P(E) = - C_0(x_0,y_0).
    \label{eq:PC}
\end{equation}

\begin{figure}[t!]
    \centering
    \includegraphics[width=0.95\linewidth]{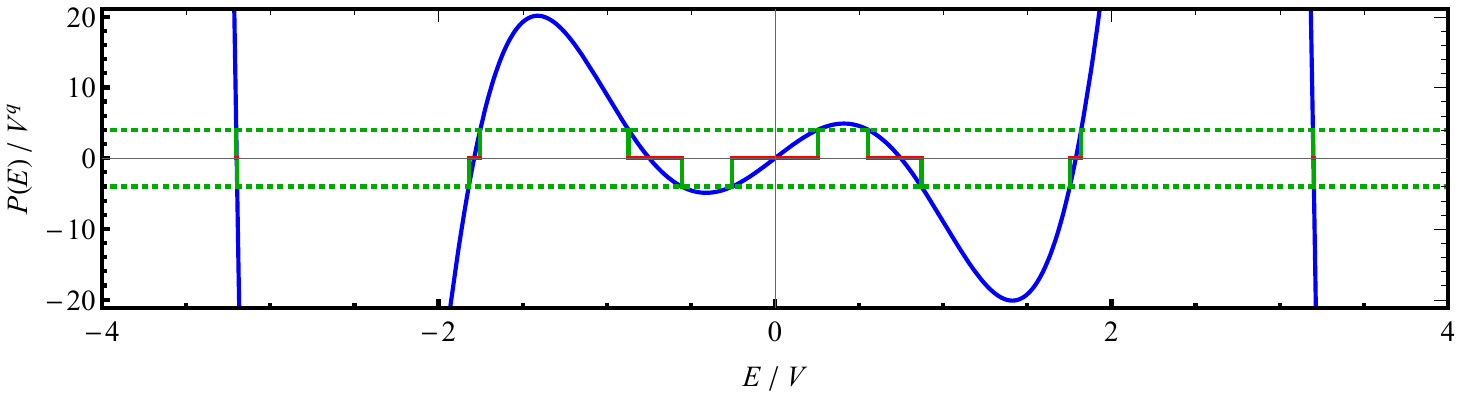}
    \caption{
    \emph{Graphical illustration of solutions to~\eqref{eq:PC} for $V_x = V_y = V$ and $\alpha = p/q = 1/7$}: $P(E)$ is shown in blue, the range of values swept out by $C_0$ is demarcated by the dashed green lines, and the corresponding range of values swept out by the roots of~\eqref{eq:PC} is marked in red on the horizontal axis.
    }
    \label{Fig:Chambers}
\end{figure}

The solutions to~\eqref{eq:PC} are plotted in Fig~\ref{Fig:Chambers} for $V_x = V_y = V$ and $\alpha = 1/7$. In this figure $P(E)$ is shown in blue, the range of values swept out by $C_0$ is demarcated by the dashed green lines, and the corresponding range of values swept out by the solutions to~\eqref{eq:PC} is marked in red on the horizontal axis. Each red interval corresponds to one band $E_j(x_0,y_0)$. Intuitively, it follows from eyeballing Fig.~\ref{Fig:Chambers} that each band takes the form
\begin{equation}
    E_j(x_0,y_0) \approx E_j^* + \frac{C_0(x_0,y_0)}{P'(E_j^*)}
\end{equation}
which may be obtained by linearizing $P(E)$ about its roots $E_j^*$. We expect the corrections to this to be small if the variation of the gradient $P'(E)$ is small over the interval in which $E_j(x_0,y_0)$ varies, i.e. to leading order, that $W_j P''(E_j^*)  \ll P'(E_j^*)$ where $W_j$ is the bandwidth. There is reason to expect this leading order analysis of the error should be expected to provide an accurate answer: $P''(E)$ varies only on the scale of the separation between successive roots, and thus we generically expect $P''(E_j^*)$ to provide a good order of magnitude estimate for $P''(E)$ for $E$ in the range $E_{j-1}^* \leq E \leq E_{j+1}^*$. In the remainder of this section, we perform such a leading order analysis to make this intuitive statement more precise.

Having obtained a form for the characteristic equation, we see that linearizing $P(E)$ about its roots $E_j^*$, yields
\begin{equation}
    0 = P'(E_j^*) (E_j(x_0,y_0) - E_j^*) + C_0(x_0,y_0) + O(P''(E_j^*)(E_j(x_0,y_0) - E_j^*)^2).
\end{equation}
The solutions to this equation give the band structure up to an error which must be estimated
\begin{equation}
    \begin{aligned}
    E_j(x_0,y_0) & = E_j^* + \frac{C_0(x_0,y_0)}{P'(E_j^*)} - O\left( \frac{C_0^2(x_0,y_0) P''(E_j^*)}{(P'(E_j^*))^3} \right)
    \\
    &
    = E_j^* - 2 V_x' \cos q x_0 - 2 V_y' \cos q y_0 - O\left( \frac{W_j^2 P''(E_j^*)}{P'(E_j^*)} \right)
    \end{aligned}
    \label{eq:ejsupp}
\end{equation}
where in the second line we have substituted~\eqref{eq:PC} and defined $V_x' = V_x^q / P'(E_j^*)$, $V_y' = V_y^q / P'(E_j^*)$ and $W_j = 4 V_x' + 4 V_y'$

Finally, we note that the ratio $P''(E_j^*)/P'(E_j^*)$ may be related to the band spacing. Specifically, for a quadratic expansion of the characteristic polynomial
\begin{equation}
    P(E) = P'(E_j^*) (E_j(x_0,y_0) - E_j^*) - C_0(x_0,y_0) + \tfrac12 P''(E_j^*)(E_j(x_0,y_0) - E_j^*)^2 + O(P'''(E_j^*)(E_j(x_0,y_0) - E_j^*)^3)
\end{equation}
this equation has roots at
\begin{equation}
    E = E_j^*, \qquad \text{and} \qquad E = E_j^* - \frac{2 P'(E_j^*)}{ P''(E_j^*) } + O \left(\frac{P'(E_j^*)^2 P'''(E_j^*)}{ P''(E_j^*)^2}\right)
\end{equation}
this yields a distance between the roots of
\begin{equation}
    \Delta_j = \frac{2 P'(E_j^*)}{ P''(E_j^*) } + O \left(\frac{P'(E_j^*)^2 P'''(E_j^*)}{ P''(E_j^*)^3}\right)
\end{equation}
which, combined with~\eqref{eq:ejsupp}, yields~\eqref{eq:Ea} in the main text.

\section{Numerical evidence for Eq.~(7)}
\label{app:89}

\begin{figure}[t!]
    \centering
    \includegraphics[width=0.8\linewidth]{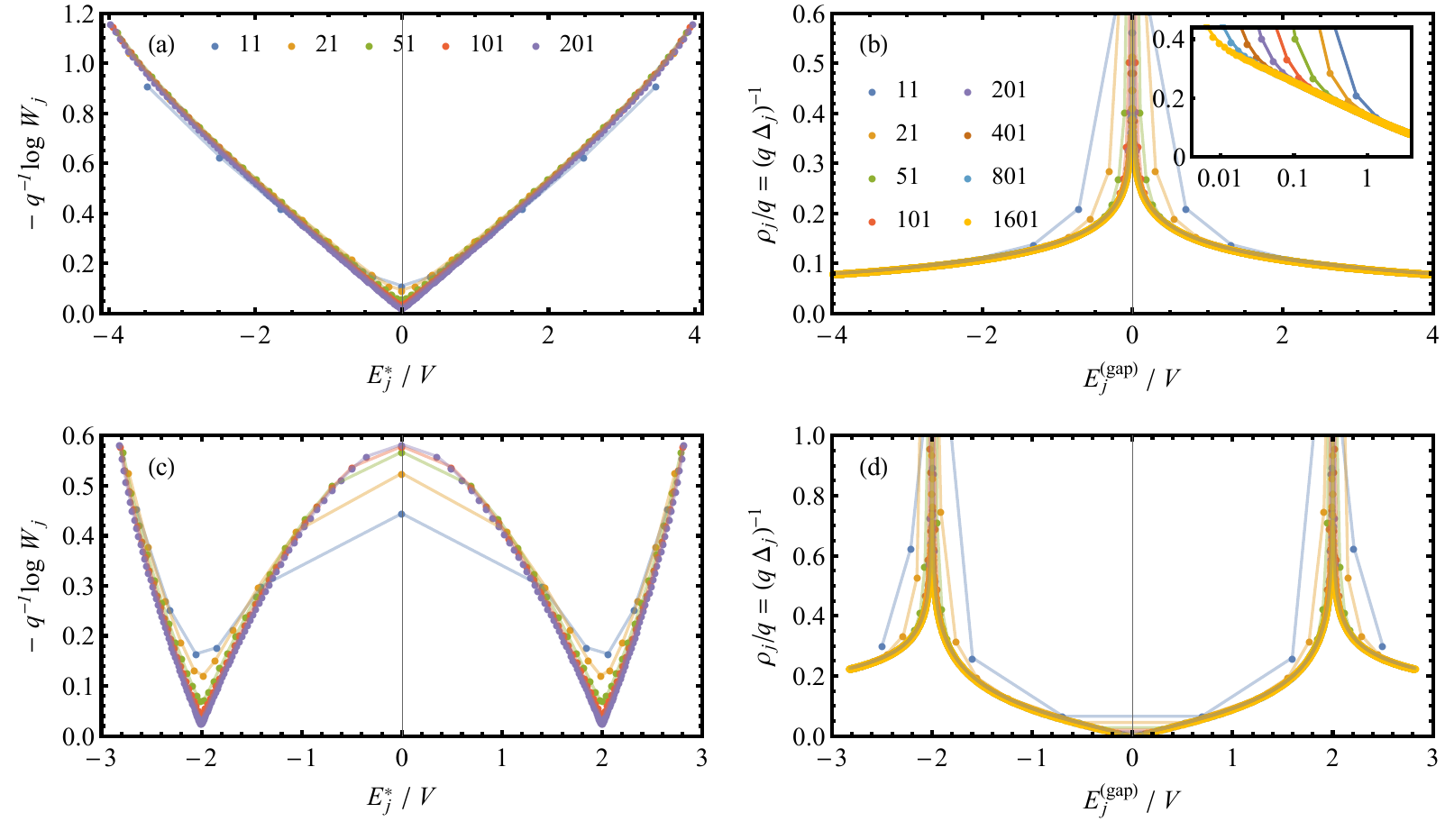}
    \caption{
    \emph{Limiting forms for the bandwidths $W_j$ and band gaps $\Delta_j$}: (a,b) the bandwidths and band gaps are analysed for $\alpha = 1/q$ in the limit $q \to \infty$ (a) $q^{-1} \log W_j$ is plotted versus $E_j^*$ for various values of $q$ (legend inset), in (b) $( \Delta_j q)^{-1}$ is plotted versus $E_j^{(\mathrm{gap})}$, the inset shows the same data on a logarithmic horizontal scale. In (c,d) analogous plots are shown for $\alpha = p/q$ with $q = 2 p + 1$.
    }
    \label{Fig:supp}
\end{figure}

Eq.~\eqref{eq:Wgap_size} is obtained via analytic arguments by Wilkinson in Refs.~\cite{wilkinson1984critical,wilkinson1987exact}. Nevertheless, here we provide some numerical evidence of this result.

Consider the HH Hamiltonian~\eqref{eq:H1} with $\alpha = 1/q$ tuned to the critical point $V_x = V_y = V$. Per the previous Appendix, this Hamiltonian has $q$ bands $E_j(x_0,y_0)$ for $j = 1 \cdots q$. We denote the extrema of each band by $E_j^{( \min )}$ and $E_j^{( \max )}$. We further denote the band centres, bandwidths, and gap centres respectively by
\begin{equation}
    E_j^* = \tfrac12 \left( E_j^{( \max )} + E_j^{( \min )} \right),
    \qquad 
    \Delta_j = \rho_j^{-1} =  E_{j+1}^{( \min )} - E_j^{( \max )} ,
    \qquad
    E_j^{(\mathrm{gap})} = \tfrac12 \left( E_{j+1}^{( \min )} + E_j^{( \max )} \right)
\end{equation}
where $\rho_j = \Delta_j^{-1}$ is the density of states at the gap centre.

In Fig.~\ref{Fig:supp}a we show the bandwidths are decaying exponentially in $q$, specifically we plot $q^{-1} \log W_j$ versus $E_j^*$ for various values of $q$ (legend inset). The different $q$ series approaches a limiting form at large $q$
\begin{equation}
    \log W_j \sim - q \ell(E_j^* / V).
\end{equation}
In Fig.~\ref{Fig:supp}b we plot $\rho_j / q$ as a function of $E_j^{(\mathrm{gap})}$, showing that in the same limit the density of states has the limiting form
\begin{equation}
    \rho_j = \Delta_j^{-1} \sim q \varrho(E_j^* / V) / V.
\end{equation}
In the inset of Fig.~\ref{Fig:supp}b the same data is shown on a log scale, showing that $\varrho(z)$ has an integrable (specifically logarithmic) divergence at $z=0$. Plots Fig.~\ref{Fig:supp}c-d show the equivalent plots in the limit of large $q$ with $q = 2 p+1$, illustrating that analogous limits occur for $\alpha \to 1/2$. Indeed similar limits apply for $\alpha$ approaching any rational.

\section{WKB analysis, and derivation of the energy rescaling RG rule Eq.~(8c)}
\label{app:WKB}

This appendix derives the energy rescaling reported in Eq.~\eqref{eq:V_scaling} in the main text, for both the lower-band and central-band decimation steps, within a single elementary framework, recovering the common prefactor $4G/\pi$. 

\paragraph*{\textbf{Prior work:}} The WKB-RG framework underlying this construction was set up by Wilkinson, first as an asymptotic scheme~\cite{wilkinson1984critical} and then in exact form~\cite{wilkinson1987exact}, where band projection of the rational $p/q$ HH model yields another HH-like model with reduced flux (see also R\"udinger and Pi\'echon~\cite{rudinger1997hofstadter} for a compact statement of the resulting flux maps). The lower-band tunnel integral was evaluated in this scheme by Szab\'o and Schneider~\cite{szabo2018non} and reduces to an integral form for the Catalan-constant previously identified by Thouless~\cite{thouless1983bandwidths}. With regard to the central band calculation, while closely related work is found in Refs.~\cite{han1994critical,helffersjostrand1988,*helffersjostrand1989,*helffersjostrand1990}, the calculation is, to our knowledge, a novel extension of this framework.

\paragraph*{\textbf{Set-up}:} The derivation is semi-classical, with $\hbar = 2\pi\alpha$ (lower band) or $\hbar_2 = 2\pi(\alpha-1/2)$ (central band) playing the role of the effective Planck constant. In each case the analysis proceeds as follows:
\begin{enumerate}
    \item Expand $H$ around reference values at which the spectrum is exactly accessible: $\alpha = n$ (the small-flux limit, relevant to the lower-band decimation) and $\alpha = n + 1/2$ (the half-integer case, relevant to the central-band decimation)
    \item Identify the relevant localized states of the reference Hamiltonian. These are harmonic oscillator orbitals around the band extrema for the lower band, Dirac zero modes around the band crossings for the central band. In each case these form a degenerate set of isolated states.
    \item Compute the WKB tunnel matrix element between neighbouring such states. Tunnelling breaks the degeneracy and broadens the previously identified states into a band, and thus determines the energy scale $V'$ of the Hamiltonian obtained by projecting into this band.
\end{enumerate}

We begin from the Hamiltonian at criticality. In this appendix we use the notation $\hat{p} \equiv \hat{y}$ and introduce the effective Planck's constant $\hbar = 2\pi\alpha$ for consistency with the usual WKB literature,
\begin{equation}
    H = H(\hat{x},\hat{p}) = -2V\cos\hat{p} - 2V\cos\hat{x}, \qquad [\hat{x},\hat{p}] = i\hbar, \quad \hbar = 2\pi\alpha.
    \label{eq:AA_app}
\end{equation}
App.~\ref{sec:hbar2npi} solves the case $\hbar = 2\pi n$ exactly. App.~\ref{sec:hbar2npi_eps} treats the neighbourhood $\hbar = 2\pi n + \epsilon$, giving the lower-band rule. Sections~\ref{sec:hbar_half} and~\ref{sec:hbar_half_eps} treat the half-integer reference $\hbar = (2n+1)\pi$ and its neighbourhood, yielding the central-band rule. App.~\ref{sec:hbar_pq} sketches the extension to general rational reference points $\alpha = p/q$.

\subsection{Case $\hbar = 2\pi n$ ($\alpha$ an integer)}
\label{sec:hbar2npi}

At $\hbar = 0$, $\hat{x}$ and $\hat{p}$ commute and $H$ is diagonal in their
joint eigenbasis $\ket{x,p}$, with spectrum
\begin{equation}
    E(x,p) = -2V\cos p - 2V\cos x,
    \label{eq:a0_energy}
\end{equation}
the energy maxima and minima are at $x,p \in \pi\mathbb{Z}$. The same spectrum applies
for all $\hbar \in 2\pi\mathbb{Z}$. This follows from
\begin{equation}
    [\e^{i\hat{p}},\e^{i\hat{x}}] = 2 i\sin(\hbar/2)\,\e^{i(\hat{x}+\hat{p})},
\end{equation}
so that the operators $\e^{i\hat{p}}$ and $\e^{i\hat{x}}$ are simultaneously
diagonalisable for $\hbar \in 2\pi\mathbb{Z}$. Using
$\e^{i\hat{p}}\ket{x} = \ket{x-\hbar}$, their joint eigenstates are the
Bloch waves
\begin{equation}
    \ket{x,p} = \sum_{k \in \mathbb{Z}} \e^{ikp}\,\ket{x + k\hbar},
\end{equation}
satisfying
\begin{equation}
    \e^{\pm i\hat{x}}\ket{x,p} = \e^{\pm ix}\ket{x,p},
    \qquad
    \e^{\pm i\hat{p}}\ket{x,p} = \e^{\pm ip}\ket{x,p},
\end{equation}
and hence diagonalising $H$ with eigenvalue $E(x,p)$ given
by~\eqref{eq:a0_energy}. The labels are redundant under
$\ket{x+\hbar,p} = \e^{-ip}\ket{x,p}$ and $\ket{x,p+2\pi}=\ket{x,p}$, so a
fundamental domain is $x\in[x_0-\hbar/2,x_0+\hbar/2]$, $p\in[-\pi,\pi]$,
on which the basis is orthonormal and complete.

\subsection{Case $\hbar = 2\pi n + \epsilon$ ($\alpha$ close to integer)}
\label{sec:hbar2npi_eps}

\subsubsection{\textbf{Harmonic oscillator expansion}}
\label{eq:HO_approx}

As observed in the previous section, for $\hbar = 0$ the extremal energies are at $x_*,p_* \in \pi \mathbb{Z}$, see Eq.~\eqref{eq:a0_energy}. For $\hbar$ small, expanding the cosines
in $H$ around any of these minima gives
\begin{equation}
    H_0 = -4V + V(\hat{x}-x_*)^2 + V(\hat{p}-p_*)^2,
    \qquad [\hat{x},\hat{p}] = i\hbar.
\end{equation}
This is a one-dimensional quantum harmonic oscillator. Writing
$\hat{x}' = \hat{x}-x_*$, $\hat{p}' = \hat{p}-p_*$ and introducing
$a = (\hat{x}'+i\hat{p}')/\sqrt{2\hbar}$, we have
\begin{equation}
    H_0 = -4V + 2\hbar V\,(a^\dagger a + \tfrac{1}{2}),
\end{equation}
with ground state energy and wavefunction
\begin{equation}
    E_0 = -4V + \hbar V,
    \qquad
    \braket{x}{\psi_0} = \frac{1}{(\pi\hbar)^{1/4}}\,\exp\!\left(-\frac{(x-x_*)^2}{2\hbar}\right).
    \label{eq:E0_HO_app}
\end{equation}
This estimate of $E_0$ becomes asymptotically accurate as $\hbar \to 0$: the wavefunction spread
\begin{equation}
    \sigma_x = \sqrt{\cexp{\hat{x}^2} - \cexp{\hat{x}}^2} = \sqrt{\hbar/2}, \qquad \sigma_p = \sqrt{\hbar/2}
\end{equation}
shrinks onto the region where $H_0$ and $H$ agree. Indeed, the first-order correction to energy is sub-leading in $\hbar$.
\begin{equation}
    \Delta E = \bra{\psi_0} (H - H_0) \ket{\psi_0}
    = 4V\big(1 - e^{-\hbar/4}\big) - V\hbar
    = -\tfrac{1}{8}V\hbar^2 + O(\hbar^3),
\end{equation}
In terms of $\alpha$, Eq.~\eqref{eq:E0_HO_app} reads
\begin{equation}
    E_0/V = -4 + 2\pi\alpha + O(\alpha^2),
\end{equation}
which is the small-$\alpha$ asymptote of the lower edge of the
Hofstadter spectrum~\cite{wilkinson1984critical,bellissard1989cstar}.

The ground states $\psi_0$ at neighbouring minima tunnel-couple, broadening the degenerate states calculated here into a band. In the lower-band decimation step this band becomes the renormalized Harper-Hofstadter model, so its width sets the energy scale $V'$. The naive estimate of the coupling, the matrix element
\begin{equation}
    t = \bra{\psi_0'} H \ket{\psi_0}
\end{equation}
with $\psi_0,\psi_0'$ oscillator ground states at neighbouring minima---e.g. $(x_*',p_*') = (x_*+2\pi,p_*)$---is quantitatively inaccurate: the Gaussian tails decay too fast in the classically forbidden region. The coupling must instead be evaluated within WKB, which we now detail.

\subsubsection{\textbf{WKB analysis}}

\paragraph*{\textbf{Framework:}} For $\hbar \ll 1$ we look for WKB-like solutions
\begin{equation}
    \braket{x}{\psi} = A(x)\exp\left(i S(x)  \right), \qquad S(x) = \frac{1}{\hbar} \int_{x_0}^x dx' \, p(x')
\end{equation}
Inserting into $H\ket{\psi} = E\ket{\psi}$ and using $\bra{x}\e^{\pm i\hat{p}}\ket{\psi} = \psi(x\pm\hbar)$, the eigenvalue equation becomes
\begin{equation}
    E = -V\left[ \frac{A(x+\hbar)}{A(x)} \exp\!\left(  \frac{i}{\hbar} \int_{x}^{x+\hbar}\!\!p(x')\,dx' \right) + \frac{A(x-\hbar)}{A(x)} \exp\!\left(  \frac{i}{\hbar} \int_{x}^{x-\hbar}\!\!p(x')\,dx' \right) \right] - 2V\cos x.
\end{equation}
Substituting the series expansions $A(x) = \sum_n \hbar^n A_n(x)$ and $p(x) = \sum_n \hbar^n p_n(x)$, the $O(\hbar^0)$ equation is the dispersion relation
\begin{equation}
    E = -2V\cos p_0(x) - 2V\cos x,
    \label{eq:dispersion_app}
\end{equation}
which determines $p_0(x)$ implicitly at given $x$ and $E$. The $O(\hbar)$ equation,
\begin{equation}
    0 = 2 p_1(x)\sin p_0(x) - i\!\left[\,2\sin p_0(x)\,\frac{A_0'(x)}{A_0(x)} + p_0'(x)\cos p_0(x)\,\right],
\end{equation}
separates into real and imaginary parts, giving
\begin{equation}
    p_1(x) = 0, \qquad A_0(x) = \frac{c}{\sqrt{\sin p_0(x)}}.
\end{equation}
Continuing to $O(\hbar^2)$, $p_1 = 0$ forces $A_1 = 0$; higher corrections will not be needed.

The dispersion~\eqref{eq:dispersion_app} admits real $p_0$ in the classically allowed region $|E + 2V\cos x| < 2|V|$. In the forbidden region $|E + 2V\cos x| > 2|V|$, $p_0$ is imaginary; writing $p_0 = i\kappa_0$ yields
\begin{equation}
    E = -2V\cosh\kappa_0(x) - 2V\cos x, \qquad A_0(x) = \frac{1}{\sqrt{\sinh\kappa_0(x)}},
    \label{eq:tunnelling}
\end{equation}
with $\kappa_1 = A_1 = 0$ as before.

\paragraph*{\textbf{Solution in the propagating region near a minimum:}} We first consider WKB solutions near an energy minimum. This recovers the harmonic-oscillator levels of App.~\ref{eq:HO_approx} in a language that we will extend to obtain the tunnel integral.

The lowest-band states are localized near minima of the modulating potential, $x_*, p_* \in 2 \pi \mathbb{Z}$. Focusing on the well at $(x_*,p_*) = (0,0)$ and expanding the cosines to quadratic order, the dispersion~\eqref{eq:dispersion_app} becomes
\begin{equation}
    E + 4V = V(x^2 + p_0^2) + \ldots
    \label{eq:harmonic_app}
\end{equation}
The classical orbits at fixed $E$ are circles of radius $r = \sqrt{E/V+4}$, enclosing action
\begin{equation}
    \oint p_0(x)\,dx = \pi r^2 = \pi(4 + E/V).
\end{equation}
EBK quantisation,
\begin{equation}
    \oint p(x)\,dx = 2\pi\hbar(n + \tfrac12),
\end{equation}
(with the $\tfrac12$ a Maslov phase from the two soft turning points), combined with $p = p_0 + O(\hbar^2)$ from the framework above, gives
\begin{equation}
    E_n = -4V + \hbar\omega(n+\tfrac12) = -4V + 4\pi\alpha V(n+\tfrac12) + O(\alpha^2)
\end{equation}
with oscillator frequency $\omega = 2V$. The ground state sits at
\begin{equation}
    E_0 = -4V + \hbar V + O(\hbar^2),
    \label{eq:E0_app}
\end{equation}
on the classical orbit $x^2 + p_0^2 = \hbar$.

\paragraph*{\textbf{Solution in the tunnelling region:}} Classically, a particle with energy $E_0$ is confined to the region $-x_\mathrm{t} < x < x_{\mathrm{t}}$, where $x_{\mathrm{t}}$ is the classical turning point at which the kinetic energy is minimised, $\cos p = 1$,
\begin{equation}
    E_0 = E(x_\mathrm{t}, 0) = -2V - 2V\cos x_{\mathrm{t}}.
\end{equation}
and combining with $E_0 = -4V + \hbar V$ from~\eqref{eq:E0_app}, the turning point lies asymptotically close to the well in the limit of small $\hbar$,
\begin{equation}
    x_{\mathrm{t}} = \sqrt{\hbar} + O(\hbar^{3/2}).
\end{equation}
To reach the adjacent well at $x_* = 2\pi$, the particle must tunnel through the forbidden region $x_{\mathrm{t}} < x < 2\pi - x_{\mathrm{t}}$. There, per~\eqref{eq:tunnelling} and~\eqref{eq:E0_app},
\begin{equation}
    \kappa_0(x) = \operatorname{arcosh}(2 - \cos x) + O(\hbar).
    \label{eq:kappa_asymp_app}
\end{equation}
In the barrier the wavefunction is exponentially small in $\int_{x_{\mathrm{t}}}^x \kappa(x')\,dx'$, and one obtains a matrix element that couples adjacent ground states from the standard WKB tunnel formula~\cite{szabo2018non,landau1981quantum}
\begin{equation}
    V' = \frac{\hbar\omega}{2\pi}\,\exp\!\left[-\frac{1}{\hbar}\!\int_{x_{\mathrm{t}}}^{2\pi - x_{\mathrm{t}}}\!\!\kappa_0(x)\,dx\right],
    \label{eq:tunnel_app}
\end{equation}
with $\omega = 2V$ the oscillator frequency and $\hbar\omega/(2\pi) = 2\alpha V$ the attempt frequency. By the symmetry $\kappa_0(2\pi - x) = \kappa_0(x)$, the action integral evaluates as
\begin{equation}
    \int_{x_{\mathrm{t}}}^{2\pi - x_{\mathrm{t}}}\!\!\kappa_0(x)\,dx
    = \int_0^{2\pi}\!\operatorname{arcosh}(2 - \cos x)\,dx + O(\hbar)
    = 8G + O(\hbar),
    \label{eq:catalan}
\end{equation}
where $G = 0.915966\ldots$ is Catalan's constant. This gives
\begin{equation}
    V' = 2V\alpha\,\exp\!\left(-\frac{4G}{\pi\alpha} + O(\alpha^0)\right),
\end{equation}
or equivalently
\begin{equation}
    \log\frac{V'}{V} \sim -\frac{4G}{\pi\alpha} = -\frac{4G\,s(\tilde\alpha)}{\pi},
\end{equation}
matching the main-text Eq.~(8c) with $s(\tilde\alpha) = 1/\tilde\alpha$ for the lower band.

\paragraph*{\textbf{Solution near a turning point:}} The WKB amplitudes $A_0 \propto 1/\sqrt{\sin p_0}$ and $A_0 \propto 1/\sqrt{\sinh \kappa_0}$ both diverge as $p_0,\kappa_0 \to 0$ at a turning point, signalling that the WKB ansatz fails there. In a neighbourhood of $x_{\mathrm{t}}$ the potential is linear to leading order, and the Schr\"odinger equation is solved exactly by Airy functions. Matching these local solutions to the WKB solutions in the allowed and forbidden regions yields the standard connection formulae~\cite{berry1972semiclassical,landau1981quantum}.

\subsection{Case $\hbar = (2n+1)\pi$ ($\alpha$ half integer)}
\label{sec:hbar_half}

We now follow a similar prescription to the $\hbar=0$ case, modified to handle the fact that $\e^{i\hat{p}}$ and $\e^{i\hat{x}}$ no longer commute. The result is a two-band model with Dirac points at $E=0$. Specifically,
\begin{equation}
    [\e^{i\hat{p}}, \e^{i\hat{x}}] = 2i\sin(\hbar/2)\,\e^{i(\hat{x}+\hat{p})} = 2i(-1)^n\,\e^{i(\hat{x}+\hat{p})} \neq 0,
\end{equation}
so $\e^{i\hat{p}}$ and $\e^{i\hat{x}}$ cannot be simultaneously diagonalised. However,
\begin{equation}
    [\e^{2i\hat{p}}, \e^{i\hat{x}}] = 2i\sin(\hbar)\,\e^{i(\hat{x}+2\hat{p})} = 0,
\end{equation}
so $\e^{2i\hat{p}}$ and $\e^{i\hat{x}}$ can be. The corresponding Bloch waves are defined on even ($+$) and odd ($-$) numbers of $\hbar$-displacements from a reference site,
\begin{equation}
    \ket{x,p,+} = \sum_{k\in\mathbb{Z}} \e^{2ikp}\,\ket{x+2k\hbar}, \qquad
    \ket{x,p,-} = \sum_{k\in\mathbb{Z}} \e^{i(2k+1)p}\,\ket{x+(2k+1)\hbar},
\end{equation}
or, in the notation of App.~\ref{sec:hbar2npi}, $\ket{x,p,+} = \ket{x,2p}$ and $\ket{x,p,-} = \ket{x+\hbar,2p}$, identifying these as a reorganisation of the integer-$\alpha$ Bloch states at doubled momentum period. They satisfy
\begin{equation}
    \ket{x+\hbar,p,\sigma} = \e^{-ip}\ket{x,p,-\sigma}, \qquad \ket{x,p+\pi,\sigma} = \sigma\ket{x,p,\sigma},
\end{equation}
so a fundamental domain is $x\in[x_0-\hbar/2,x_0+\hbar/2]$, $p\in[-\pi/2,\pi/2]$, on which the basis is orthonormal and complete. Acting with $\e^{\pm i\hat{p}}$ and $\e^{\pm i\hat{x}}$,
\begin{equation}
    \begin{aligned}
        \e^{\pm i\hat{p}}\ket{x,p,\sigma} &= \e^{\pm ip}\,\sigma_x\ket{x,p,\sigma}, \\
        \e^{\pm i\hat{x}}\ket{x,p,\sigma} &= \e^{\pm ix}\,\sigma_z\ket{x,p,\sigma},
    \end{aligned}
\end{equation}
with the Pauli matrices acting on the $\sigma$ indices. The Hamiltonian thus acts as
\begin{equation}
    H\ket{x,p,\sigma} = H_2(x,p)\ket{x,p,\sigma}, \qquad H_2(x,p) = -2V\cos(p)\,\sigma_x - 2V\cos(x)\,\sigma_z,
\end{equation}
which diagonalises to bands
\begin{equation}
    E_\pm(x,p) = \pm 2V\sqrt{\cos^2 p + \cos^2 x}.
\end{equation}
The two bands touch in a Dirac cone at $(x,p) = (\pi/2,\pi/2)$. This zero-energy crossing seeds the central-band states analysed in the next subsection, where a small detuning $\hbar \to (2n+1)\pi + \epsilon$ opens a gap and localizes them.

\subsection{Case $\hbar = (2n+1)\pi + \epsilon$ ($\alpha$ close to half integer)}
\label{sec:hbar_half_eps}

We now follow a similar prescription to the $\hbar = 2\pi n + \epsilon$ case, modified for the two-band structure. Where previously we had a single band with orbitals built from a Harmonic oscillator approximation, we now have a two-band model with localized states built from a Dirac-point expansion.

\subsubsection{\textbf{Dirac point expansion}}
\label{sec:DPE}

\paragraph*{\textbf{Isospectral Hamiltonian}:}
Exploiting the structure identified in the previous section away from the point $\hbar=(2n+1)\pi$, we consider the Hamiltonian
\begin{equation}
    H_2 = H_2(\hat{x}_2,\hat{p}_2) = - 2 V \cos (\hat{p}_2)  \sigma_x- 2 V \cos (\hat{x}_2)  \sigma_z.
\end{equation}
We first show that for the choice
\begin{equation}
    [\hat{x}_2, \hat{p}_2] = i \hbar_2 \qquad \text{with} \qquad \hbar_2 = \hbar - \pi,
\end{equation}
the Hamiltonian $H_2$ has the same spectrum as $H$. To see this we consider the action of $H$ on the basis $\ket{x}$, and the action of $H_2$ on the basis $\ket{x_{0},\sigma_0}$, the simultaneous eigenbasis of $\hat{x}_2$ and $\sigma_z$. In the former case, setting $\ket{n} = \ket{x_0+n \hbar}$, the eigenvalue equation reads
\begin{equation}
    (H-E) \ket{n} = -V\ket{n+1}-V\ket{n-1}-[E+2 V \cos(x_0 + \hbar n)] \ket{n};
    \label{eq:tdse_1}
\end{equation}
whereas, identifying $\ket{2n} = \ket{x_0'+2n \hbar_2 ,+}$ and $\ket{2n+1} = \ket{x_0'+(2n+1) \hbar_2 ,-}$,
\begin{equation}
    \begin{aligned}
        (H_2-E) \ket{n} & = -V \ket{n-1}- V \ket{n+1}
        - [E+ 2 V (-1)^n \cos(x_0' + n \hbar_2)]\ket{n}
        \\
        &= -V \ket{n-1}- V \ket{n+1}
        - [E+ 2 V \cos(x_0' + n (\hbar_2 + \pi) )]\ket{n}.
    \end{aligned}
    \label{eq:tdse_2}
\end{equation}
The two eigenvalue problems~\eqref{eq:tdse_1} and~\eqref{eq:tdse_2} are identical if $x_0' = x_0$ and $\hbar_2 = \hbar - \pi$. The same construction applies for $\sigma_0=-1$ on identifying $\ket{2n} = \ket{x_0'+2n \hbar_2 ,-}$ and $\ket{2n+1} = \ket{x_0'+(2n+1) \hbar_2 ,+}$, with the only change being $x_0' = x_0 - \pi$.

This establishes that $H_2$ and $H$ are isospectral, with the action of $H_2$ on the lattice generated by translations of $\ket{x_2,\sigma_0}$ by $\hbar_2$ matching the action of $H$ on the lattice generated by translations of $\ket{x_0}$ by $\hbar$. This is weaker than unitary equivalence of the full operators, which would further require the spectral multiplicities of $H$ and $H_2$ to match, but is sufficient in the present context.

\paragraph*{\textbf{Zero energy modes}:} We can then study the low energy physics by expanding around the Dirac point. Shifting the Dirac point to the origin via $\hat{x}_2 \to \hat{x}_2+\pi/2$, $\hat{p}_2 \to \hat{p}_2+\pi/2$, sending $\sigma_z \to \sigma_y$ (a unitary transformation), and expanding in small $\hat{p}_2$, $\hat{x}_2$,
\begin{equation}
    \begin{aligned}
        H_2' &= UH_2 U^\dagger
        \\
        & =  2 V \sin (\hat{x}_2)  \sigma_y+ 2 V \sin (\hat{p}_2)  \sigma_x
        \\
        & \approx 2 V \hat{x}_2  \sigma_y+ 2 V \hat{p}_2\sigma_x.
    \end{aligned}
\end{equation}
This Hamiltonian is solved by the usual process: defining creation and annihilation operators
\begin{equation}
    a^\dagger = \frac{\hat{p}_2 + i \hat{x}_2}{\sqrt{2 \hbar_2}},
\end{equation}
we obtain
\begin{equation}
    H_2' \approx 2V\sqrt{2 \hbar_2}\left( a^\dagger \sigma_- + a \sigma_+ \right),
\end{equation}
with eigenstates and eigenvalues
\begin{equation}
    \ket{\psi_n^\pm} = \frac{1}{\sqrt{2}}\binom{\ket{n-1}}{\pm \ket{n}}, \qquad 
    E_{n}^\pm = \pm 2 V \sqrt{2 \hbar_2 n}
\end{equation}
for integer $n>0$ with $\ket{n}$ the usual number states, while for $n=0$
\begin{equation}
    \ket{\psi_0} =  \ket{0,-}, \qquad 
    E_{0} = 0.
\end{equation}
The zero mode is protected from being shifted by higher-order corrections by the chiral symmetry
\begin{equation}
    \{ H_2' , \sigma_z \} = 0.
\end{equation}
Furthermore, the zero-energy state has position and momentum spread
\begin{equation}
    \sigma_{x_2} = \sqrt{\hbar_2/2}, \qquad \sigma_{p_2} = \sqrt{\hbar_2/2},
\end{equation}
so for $\hbar_2 \ll 1$ the state is localized on a scale much smaller than that on which $H_2'$ varies, validating the small angle expansion.

$H_2'$ has similar localized zero-energy states in the vicinity of all positions $x_2 \in \pi\mathbb{Z}$. These states tunnel-couple and broaden into the central band.

\subsubsection{\textbf{WKB analysis}:}
For $\hbar_2 \ll 1$ we look for solutions of the form
\begin{equation}
    \braket{x,\sigma}{\psi} = \phi(x,\sigma)\exp\left( \frac{i}{\hbar_2} \int_{x_0}^x dx' \, p(x') \right),
\end{equation}
for which the action of the Hamiltonian is
\begin{equation}
     \bra{x,\sigma} H_2 \ket{\psi} = -V \Bigg[\, \psi_{-\sigma}(x+\hbar_2)\exp\!\left(\frac{i}{\hbar_2}\int_x^{x+\hbar_2}p(x')\,dx'\right)+ \psi_{-\sigma}(x-\hbar_2)\exp\!\left(\frac{i}{\hbar_2}\int_x^{x-\hbar_2}p(x')\,dx'\right) \Bigg] - 2V\sigma\cos(x)\psi_\sigma(x).
\end{equation}
Assuming an expansion
\begin{equation}
    \phi(x,\sigma) = \sum_n \hbar_2^n \phi_{\sigma,n}(x), \qquad p(x) = \sum_n \hbar_2^n p_n(x),
\end{equation}
expanding to leading order yields the Schr\"odinger equation
\begin{equation}
    \Big(H_2(x,p_0(x)) - E \Big)\binom{\phi_+(x)}{\phi_-(x)} = 0,
\end{equation}
with energies
\begin{equation}
    E_\pm(x,p_0(x)) = \pm 2 V\sqrt{\cos^2 p_0(x) + \cos^2 x},
    \label{eq:e_2H}
\end{equation}
an implicit equation for $p_0(x)$ given $x$ and energy $E$.

For the $E=0$ mode, the classically allowed region degenerates to the isolated Dirac points, and the usual WKB framework with extended classical orbits does not directly apply. However, we have already shown that the zero modes exist, and all that remains to calculate in WKB is the tunnelling integral that broadens them into a band. In the tunnelling regions $p$ is complex, and we write $p = \pi/2 + i\kappa$. For the $E=0$ mode,
\begin{equation}
    \kappa_0 = \operatorname{arcsinh}|\cos x|,
\end{equation}
so the real part of the tunnelling action between the Dirac zero modes at $(x,p) = (-\pi/2,\pi/2)$ and $(\pi/2,\pi/2)$ is
\begin{equation}
    \begin{aligned}
        \frac{1}{\hbar_2} S & = -\frac{1}{\hbar_2}\int_{-\pi/2}^{\pi/2} dx \, \kappa_0(x) + O(\hbar_2^0)
        \\
        & = -\frac{1}{\hbar_2}\int_{-\pi/2}^{\pi/2} dx \, \operatorname{arcsinh} |\cos x| + O(\hbar_2^0)
        \\
        & = -\frac{2 G}{\hbar_2} + O(\hbar_2^0),
    \end{aligned}
\end{equation}
where, as in~\eqref{eq:catalan}, $G$ is Catalan's constant. Consequently, projecting into the central band,
\begin{equation}
    -\log \left|\frac{V'}{V}\right| \sim \frac{2 G}{\hbar_2} = \frac{2G}{2\pi|\alpha-1/2|} = \frac{4G}{2\pi(1-2\tilde{\alpha})} \sim \frac{4 G}{\pi}s(\tilde{\alpha}),
\end{equation}
where $s(\tilde\alpha) = \tilde\alpha/(1-2\tilde\alpha)$ for the central band and $\tilde\alpha = \min(\alpha,1-\alpha)$ for $\alpha\in[0,1]$, as in the main text.

\subsection{Extension to other $\alpha$}
\label{sec:hbar_pq}

As a final comment regarding the derivation of the RG rules for Harper Hofstadter model, we note that the construction used here may be extended to the neighbourhood of other $\alpha = p/q$ for coprime $p,q$. Specifically, one may show that
\begin{equation}
    H_q = - V \left( \e^{i \hat{p}_q} X_q + \mathrm{h.c}\right) - V \left( \e^{i \hat{x}_q}Z_q^p + \mathrm{h.c} \right)
\end{equation}
where $X_q$ and $Z_q$ are the $q \times q$ clock and shift matrices
\begin{equation}
    Z_q X_q = \e^{2\pi i/q} X_q Z_q
\end{equation}
and 
\begin{equation}
    [\hat{x}_q,\hat{p}_q] = i \hbar_q
\end{equation}
is isospectral to the HH model with
\begin{equation}
    \hbar = \hbar_q + 2\pi p / q.
\end{equation}
allowing similar WKB analyses of these models for small $\hbar_q$.

\subsection{Extension to generalised Harper-Hofstadter models}

Finally we comment on how this analysis may be extended to a generalised Harper-Hofstadter models, specifically models of the form
\begin{equation}
    H_\mathrm{G} = V(\hat{x},\hat{p}) = \sum_{n,m\in \mathbb{Z}} V_{nm} \e^{ i (n \hat{x} + m \hat{p})}
\end{equation}
where $V(x,p)$ is a smooth periodic function with no open equipotentials over a finite interval of energies. This situation is forced for example if the potential has $C_4$ a symmetry centre $V(x,p)=V(p,-x)$ (as in the HH model) in accordance with the following observations (i) in a periodic potential, all open equipotentials must run parallel to other equipotentials related by lattice vectors (ii) $C_4$ maps open equipotentials to open equipotentials that run perpendicular (ii) it is perturbatively unstable for open equipotentials to cross.

\paragraph*{\textbf{Lower band RG}:} The lower band RG rule generalises straightforwardly. One finds the minima $x_*,p_*$ corresponding to the minima of $V(x,p)$. For sufficiently small $\hbar$, a quadratic expansion about this point yields harmonic oscillator states. Calculating the tunnelling integral between these states one generically finds tunnelling integrals, and hence energy renormalization factors, that are exponentially small in $1/\hbar$. To see this, consider the tunnelling integral coupling two extrema $(x_*,p_*) \to (x_*',p_*')$. The tunnelling integral is dominated by the instanton trajectory $\gamma$ in the $x,p$ plane which minimises the action between $(x_*,p_*) \to (x_*',p_*')$. Along $\gamma$ the classical equation $V(x,p) = V(x_*, p_*)$ has imaginary momentum $p = i\kappa(x)$. We find
\begin{equation}
    \log \frac{V'}{V} \sim -\frac{1}{\hbar}\, S, \qquad S = \int_\gamma \kappa(x)\, dx
\end{equation}
where $\gamma$ is the saddle point path connecting the minima in the $(x,p)$ plane, and $\kappa(x)$ is determined along the saddle point trajectory $\gamma$ by
\begin{equation}
    V(x_*, p_*) = V(x, i\kappa(x)).
\end{equation}
Consequently one finds
\begin{equation}
    \log \frac{V'}{V} \sim -\frac{c}{\alpha}
\end{equation}
where $c$ is a model dependent quantity.

\paragraph*{\textbf{Central band RG}:} The central band RG rule also proceeds essentially identically. Here we do not attempt to provide the most general treatment, but sketch the calculation for a broad class of models, showing the central band projection yields $V'$ which is exponentially small in $1/\hbar_2$ as in the HH case.

Starting with $\hbar = 2\pi(k+1/2)$ for integer $k$, we write $H_\mathrm{G}$ in the $\ket{x,p,\sigma}$ basis. Using the relation
\begin{equation}
    e^{i(n \hat{x} + m \hat{p})} \ket{x, p, \sigma} = e^{i nm \hbar / 2}\, e^{i(nx + mp)}\, \sigma_x^n\, \sigma_z^m\, \ket{x, p, \sigma}.
\end{equation}
we obtain
\begin{equation}
    H_\mathrm{G} \ket{x,p,\sigma} = H_{2}(x,p)\ket{x,p,\sigma}
\end{equation}
where $H_2$ may be written as
\begin{equation}
        H_2(x,p) =h_0(x,p) + \vec{h}(x,p)\cdot\vec{\sigma}
\end{equation}
with the coefficients given by 
\begin{equation}
    \begin{aligned}
        h_0(x,p) &= \tfrac{1}{4}\big[\,V(x,p) + V(x{+}\pi, p) + V(x, p{+}\pi) + V(x{+}\pi, p{+}\pi)\,\big] \\[4pt]
        h_1(x,p) &= \tfrac{1}{4}\big[\,V(x{+}\tfrac{\pi}{2}, p) - V(x{+}\tfrac{\pi}{2}, p{+}\pi) + V(x{-}\tfrac{\pi}{2}, p) - V(x{-}\tfrac{\pi}{2}, p{+}\pi)\,\big] \\[4pt]
        h_2(x,p) &= \tfrac{1}{4}\big[\,V(x{+}\tfrac{\pi}{2}, p{+}\tfrac{\pi}{2}) - V(x{+}\tfrac{\pi}{2}, p{-}\tfrac{\pi}{2}) - V(x{-}\tfrac{\pi}{2}, p{+}\tfrac{\pi}{2}) + V(x{-}\tfrac{\pi}{2}, p{-}\tfrac{\pi}{2})\,\big] \\[4pt]
        h_3(x,p) &= \tfrac{1}{4}\big[\,V(x, p{+}\tfrac{\pi}{2}) - V(x{+}\pi, p{+}\tfrac{\pi}{2}) + V(x, p{-}\tfrac{\pi}{2}) - V(x{+}\pi, p{-}\tfrac{\pi}{2})\,\big]
\end{aligned}
\end{equation}
Diagonalising the $2 \times 2$ matrix $H_2$ yields energies 
\begin{equation}
    E_{\pm}(x,p) = h_0(x,p) \pm |\vec{h}(x,p)|
    \label{eq:epm}
\end{equation}
and hence a gap $|E_+ - E_-|=2|\vec{h}(x,p)|$. We denote the minimum of this gap as $(x_*,p_*)$. 

Following the same framework as Sec.~\ref{sec:DPE}, we extend to $\hbar = 2\pi (k+1/2) + \hbar_2$. We expand this point to leading order with the replacement $\hat{x}-x_* \to \hat{x}_2$, $\hat{p}-p_* \to \hat{p}_2$  hamiltonian
\begin{equation}   
    \begin{aligned}
        H & \approx \sum_{\mu=0}^3 \sigma_\mu (a_\mu \hat{p}_2 + b_\mu \hat{x}_2 + c_\mu ) 
    \end{aligned}
    \label{eg:tilted}
\end{equation}
where $[\hat{x}_2,\hat{p}_2] = i \hbar_2$ and
\begin{equation}
    a_\mu = \left.\partial_p h_\mu(x_*,p)\right|_{p=p_*}, \qquad b_\mu = \left.\partial_x h_\mu(x,p_*)\right|_{x=x_*}, \qquad c_\mu = h_\mu(x_*,p_*)
\end{equation}
The condition that $(x_*,p_*)$ corresponds to a gap minimum enforces the conditions
\begin{equation}
    \vec{a}\cdot\vec{c} = \vec{b}\cdot\vec{c} = 0
\end{equation}
Eq.~\eqref{eg:tilted} is a massive, tilted, anisotropic Dirac Hamiltonian. 

The properties of a tilted Dirac hamiltonian depend discretely on the extent of the tilt, here for simplicity we restrict to sub-critical tilting (i.e. type I Dirac cones). To unpack this point: consider the canonical Dirac Hamiltonian, one cone of energies extends to positive energies $E>0$, and one to negative energies $E<0$. In contrast, for a $90^{\circ}$ tilted Dirac hamiltonian both cones extend infinitely far to both positive and negative energies. For general tilts a (massive) Dirac cone is considered type I if from some energy $E_0$ one cone lies entirely in $E > E_0$ and the other $E < E_0$. We restrict to the type I case, which holds provided
\begin{equation}
    \delta = |\vec{a} \times \vec{b}|^2 - |a_0 \vec{b} - b_0 \vec{a}|^2 > 0.
\end{equation}
The massive, tilted, anisotropic, type I Dirac Hamiltonian is exactly solvable and with the same quantum number $n$~\cite{tchoumakov2016magnetic} as in the canonical case. Its eigenstate energies are given by
\begin{equation}
    \begin{aligned}
        E_0 = c_0 - s_0\eta |\vec{c}|,
        \qquad
        E_{n>0}^\pm = c_0\pm\eta\sqrt{|\vec{c}|^2+2n \eta |\hbar_2||\vec{a}\times\vec{b}| }
    \end{aligned}
\end{equation}
where $s_0$ sets the sign of the zero mode, and $\eta$ quantifies the suppression of the gap due to the tilt
\begin{equation}
    s_0= \operatorname{sign}\left(\hbar_2[\vec{c}\cdot(\vec{a}\times\vec{b})]\right), \qquad  \eta = \frac{\sqrt{\delta}}{|\vec{a}\times\vec{b}|}.
\end{equation}

Similar $n=0$ modes will occur at all $(x_* + 2\pi n,p_*+2\pi m)$. We identify this family of states with the generalisation of the HH central band. The corresponding bandwidth can be calculated via the tunnelling integral coupling these states in direct generalisation of the WKB methods used in this section. This will result in a renormalized bandwidth $V'$ that is exponentially small in $1/\hbar_2$
\begin{equation}
    \log \frac{V'}{V} \sim -\frac{1}{\hbar_2}\, S, \qquad S = \int_\gamma \kappa(x)\, dx
\end{equation}
as found in the HH case, where the action in the tunnelling region is determined by the distance in energy from the nearest states in the usual way
\begin{equation}
    E_\sigma(x,i\kappa_0(x)) = E_0
\end{equation}
with $E_\sigma$ as in~\eqref{eq:epm}, with both $\sigma$ and the path $\gamma$ chosen to minimise the action.

\section{Properties of the folded Gauss map $B(\talpha)$}
\label{app:T}

In this section we demonstrate several properties of the folded Gauss map $B:[0,1/2]\to[0,1/2]$ corresponding to an RG where we project into the lower band at each step
\begin{equation}
    B(\talpha) = \min\left[ b(\talpha) , 1 - b(\talpha)\right], \qquad b(\talpha) = \frac{1}{\talpha} - \left\lfloor \frac{1}{\talpha} \right\rfloor,
\end{equation}
as defined in the main text in Eq.~\eqref{eq:beta_scaling}. Specifically we calculate the unique invariant measure of $B$, and further show that $B$ is (i) chaotic, and calculate the Lyapunov exponent $\Lambda$, (ii) ergodic, (iii) gapped, and calculate the spectral gap $\Delta$, (iv) Gibbs-Markov. 

Results for an RG projecting into the middle band at each step, as used in the latter part of the paper, follow by the same methods.

Consider an initial set values $\talpha_0^{(i)} \in [0,1/2]$ characterized by a smooth distribution $g_0(\talpha)$. Each of these values can be renormalized to yield $\talpha_n^{(i)} = B^n(\talpha_0^{(i)})$, which is also characterised by a smooth distribution function $g_n(\talpha)$. As $n$ is taken large $g_n$ converges to the unique steady state distribution
\begin{equation}
    g_n(\talpha) \to f(\talpha) = \frac{1}{\log \varphi} \cdot \frac{\varphi^3}{\varphi^3 + \talpha-\talpha^2}
    \label{eq:f_app}
\end{equation}
where $\varphi = (1 + \sqrt{5})/2$ is the golden ratio. Moreover, the deviation from the limiting distribution is exponentially small in $n$
\begin{equation}
    \log |g_n(\talpha) - f(\talpha)| \sim  - n\Delta, \qquad \text{where} \qquad \Delta = 3.7856665519818449128 \ldots
    \label{eq:Delta}
\end{equation}
is the spectral gap of $B$. The statement~\eqref{eq:Delta} together with $\Delta>0$ demonstrates the ergodicity of the map $B$. Moreover, the large value $\Delta \gg 1$, indicates that the convergence of $g_n(x)$ to $f(x)$ occurs rapidly over an $O(1)$ number of steps. Eq.~\eqref{eq:Delta} is the main result of this section. Prior to the main result, we arrive at two further results: (i) we show that $f(\talpha)$ in~\eqref{eq:f_app} is the steady state, and (ii) we show that the map $B$ is chaotic with maximal Lyapunov exponent
\begin{equation}
    \Lambda = \frac{\pi^2}{6 \log \varphi}.
\end{equation}

\subsection{Steady state distribution of $B$}

The sequence of distributions $g_n$ are defined by recursive application of the map $B$, i.e. $g_{n+1}(\talpha) = [B g_n](\talpha)$, where the action of $B$ on $g$ defines the associated Perron–Frobenius operator, given explicitly by
\begin{equation}
    \begin{aligned}
     {[B g](\talpha)} & : = \int_0^{1/2} d\talpha' \delta(\talpha - B(\talpha')) g(\talpha')  = \sum_{q=2}^\infty \left[ \frac{g\left(\frac{1}{q+\talpha}\right)}{(q+\talpha)^2} + \frac{g\left(\frac{1}{q+1-\talpha}\right)}{(q+1-\talpha)^2}\right].
     \end{aligned}
\end{equation}
Note that the action of $B$ on the space of distributions $g$ is linear, $[B(g+h)] = [B g] + [B h]$, and thus the steady state distribution $f$ is obtained as the leading eigenfunction of $B$, which has a corresponding eigenvalue of unity
\begin{equation}
    [B f](\talpha) = f(\talpha).
\end{equation}
It is then straightforward to verify that~\eqref{eq:f_app} satisfies this relation. The uniqueness of this solution is verified numerically in App.~\ref{app:ergB}.

\subsection{Chaoticity of $B$}

The Lyapunov exponent of a discrete map is given by
\begin{equation}
    \Lambda = \lim_{n \to \infty} \frac{1}{n} \sum_{m = 1}^n \log |B'(\talpha_n)|
\end{equation}
where $\talpha_n = B^n(\talpha_0)$, $B'(\talpha)$ is the derivative of $B(\talpha)$ and the Lyapunov exponent $\Lambda$ is independent of $\talpha_0$ due to the ergodicity of $B$.

Moreover, as $B$ is ergodic, $\Lambda$ may be straightforwardly evaluated using the steady state distribution $f(\talpha)$
\begin{equation}
\begin{aligned}
    \Lambda & = \int_0^{1/2}d \talpha \, f(\talpha) \, \log| B(\talpha)|
    = - 2 \int_0^{1/2}d \talpha \, f(\talpha) \, \log \talpha 
    = \frac{\pi^2}{6 \log \varphi}
    \end{aligned}
    \label{eq:lyapunov_app}
\end{equation}
In Eq.~\eqref{eq:lyapunov_app} we have used that $| B(\talpha)| = \talpha^{-2}$ except at a measure zero set of points, where the derivative is undefined. As $\Lambda > 0$, $B$ is chaotic.

\subsection{Ergodicity and spectral gap of $B$}
\label{app:ergB}

As $B$ is a linear operator, it has a spectrum of eigenvalues $\beta_k \geq 0$ with associated eigenfunctions $v_k(\talpha)$ which form a complete basis
\begin{equation}
    [B v_k](\talpha) = \beta_k v_k(\talpha).
    \label{eq:B_eigs}
\end{equation}
In principle the spectrum of eigenvalues may have discrete and continuous components, though in the present case we find only a discrete spectrum allowing us to index them in descending order $1 = |\beta_0| \geq |\beta_1| \geq |\beta_2| \geq \cdots$. The distribution at late times is found by projecting onto the subspace of eigenfunctions with eigenvalues $|\beta_k| =1$. If there is exactly one such eigenvalue, which we denote as $\beta_0 = 1$ (the eigenvalue cannot have a phase as $g_n(\talpha)$ is strictly non-negative), then the steady state distribution $f(\talpha)$ is unique, independent of $g_0$, and given by the corresponding eigenfunction $f = v_0$. The deviation of $g_n$ from $v_0$ is then determined by the first sub-leading eigenvalue: $|f - g_n| = O(\beta_1^n) = O(\e^{- \Delta n})$, where we have defined
\begin{equation}
    \Delta = - \log |\beta_1|
\end{equation}
as the spectral gap of $B$. We employ a numerical approach previously used in Ref.~\cite{briggs2003precise,flajolet1995gauss} to calculate the spectral gap of the Gauss map. 

The eigenfucntion(s) $v_k(\talpha)$ may be obtained as the solutions to the eigenvalue equation~\eqref{eq:B_eigs}, however in the absence of an analytic technique to solve this equation, we resort to numerics. To numerically tackle this problem we first re-write in terms of the coordinate $y = 1/2 - \talpha \in [0,1/2]$. In this coordinate $B$ has the action
\begin{equation}
     [B g](y) = \sum_{h} \left[ \frac{g\left(\tfrac12 - \frac{1}{h - y}\right)}{(h- y)^2} + \frac{g\left(\tfrac12 - \frac{1}{h+y }\right)}{(h+y)^2}\right]
\end{equation}
where the sum is taken over the half-integers $h = \tfrac52, \tfrac72, \tfrac92, \tfrac{11}{2} \ldots$. To make the problem numerically tractable we subsequently write $B$ in a basis spanned by a countable set of basis elements. For simplicity we choose the basis monomials
\begin{equation}
    u_p(y) = y^p = (1/2 - \talpha)^p
\end{equation}
upon which $B$ acts as
\begin{equation}
    \begin{aligned}
     [B u_p](y) &= \sum_{h} \frac{1}{2^p}\left[ \frac{\left(1 - \frac{2}{h - y}\right)^p}{(h- y)^2} + \frac{\left(1 - \frac{2}{h+y }\right)^p}{(h+y)^2}\right]
     \\
     & = \sum_{h} \frac{1}{2^{p+2}} \sum_{k = 0}^p \binom{p}{k}\left( -\frac{2}{h}\right)^{k+2}\left[ \left( \frac{1}{1 - y/h}\right)^{k+2} + \left( \frac{1}{1+y/h }\right)^{k+2}\right]
     \\
     & = \sum_{h} \frac{1}{2^{p+2}} \sum_{k = 0}^p \binom{p}{k}\left( -\frac{2}{h}\right)^{k+2}\left[ 2 \sum_{n = 0}^\infty \binom{2n + k + 1 }{2n} \left(\frac{y}{h}\right)^{2n}
     \right]
     \end{aligned}
\end{equation}
Recalling the definition of the Hurwitz zeta function $\zeta(s,a) = \sum_{n=0}^\infty (n+a)^{-s}$, and rearranging we find
\begin{equation}
    [B u_p](y) =  \sum_{k = 0}^p \sum_{n = 0}^\infty  \binom{p}{k} \binom{2n + k + 1 }{2n}   
    \frac{(-1)^k}{2^{p - k - 1}} \zeta(2n + k + 2,5/2) u_{2n}(y).
    \label{eq:Tup}
\end{equation}
We re-write~\eqref{eq:Tup} to define $M_{pq}$, the transfer matrix on the basis on monomials $u_p$ we 
\begin{equation}
    [B u_p](y) = \sum_{q = 0}^\infty M_{pq} u_q(y)
\end{equation}
where the matrix elements are given by 
\begin{equation}
    M_{pq} = \begin{cases}
    \displaystyle
    \frac{1}{2^{p-1}}\sum_{k = 0}^p  \binom{p}{k} \binom{q + k + 1 }{q}  
    (-2)^k\zeta(q + k + 2,5/2) & \qquad (q  \text{\,\, even}),
    \\[15pt]
    0 & \qquad (q  \text{\,\, odd}).
    \end{cases}
\end{equation}
Finally, we note $M_{pq}$ can be brought to a compact numerically stable integral form 
\begin{equation}
    M_{pq} = \frac{1}{q!\,2^{p+q+2}}\int_0^\infty \frac{u^q\, e^{-u}}{\sinh(u/4)}\,\bigl[L_p(u) - L_{p+1}(u)\bigr]\,du \qquad \quad (q\text{ even})
    \label{eg:m_stable}
\end{equation}
Eq.~\eqref{eg:m_stable} is obtained by inserting the integral representation
\begin{equation}
    \zeta(s,5/2) = \frac{1}{\Gamma(s)}\int_0^\infty \frac{t^{s-1}e^{-5t/2}}{1-e^{-t}}\,dt
\end{equation}
into the sum and using $\binom{q+k+1}{q}/(q+k+1)! = 1/(q!\,(k+1)!)$ moves all $k$-dependence into the sum
\begin{equation}
    S_p(t) \equiv \sum_{k=0}^p \binom{p}{k}\frac{(-2t)^k}{(k+1)!}.
\end{equation}
This sum can be evaluated in closed form. Writing $1/(k+1)! = \int_0^1 u^k/k!\, du$ and using the Laguerre polynomial definition $L_p(x) = \sum_k \binom{p}{k}(-x)^k/k!$ gives $S_p(t) = \int_0^1 L_p(2tu)\,du$, which by $\int_0^x L_n(t')\,dt' = L_n(x) - L_{n+1}(x)$ evaluates to
\begin{equation}
    S_p(t) = \frac{L_p(2t) - L_{p+1}(2t)}{2t}.
\end{equation}
Substituting $u = 2t$ and simplifying the prefactor via $e^{-5u/4}/(1-e^{-u/2}) = e^{-u}/(2\sinh(u/4))$ yields the integral form~\eqref{eg:m_stable}.

The spectrum of $M$, and hence $B$, may then be numerically estimated by evaluating $M_{pq}$ up to a cut-off $p,q \leq p_{\max}$ and diagonalising. The eigenvalues $\beta_k$ are found to be discrete, non-degenerate and exponentially decaying in $k$. As a result the values of low order eigenvalues converge exponentially as a function of $p_{\max}$, allowing them to be accurately numerically estimated. The numerical limitation is the evaluation of the matrix elements, which require high precision numerics for even moderately large $p_{\max}$. Numerically extracted values for the magnitudes of the first five sub-leading eigenvalues are given below (to 20 significant figures)
\begin{equation}
    \begin{aligned}
       - \log |\beta_0| &= 0
     \\
     \Delta = \Delta_1 = - \log |\beta_1| & = 3.7856665519818449128
     \\
     \Delta_2 = - \log |\beta_2| & = 6.7251453074741971174
     \\
     \Delta_3 = - \log |\beta_3| & = 11.339665867968595165
     \\
     \Delta_4 = - \log |\beta_4| & = 12.043871233196576668
     \\
     \Delta_5 = - \log |\beta_5| & = 16.966376007200018885
    \end{aligned}
\end{equation}

Indeed, as expected, the associated leading eigenfunction is found to be
\begin{equation}
    f(\talpha) \propto \sum_{q = 0}^\infty \left( \frac{ 1 - 2 \talpha}{\varphi^3} \right)^{2q}
    \label{eq:f_sum_app}
\end{equation}
where $\varphi = (1 + \sqrt{5})/2$ is the Golden Ratio. Performing the sum in~\eqref{eq:f_sum_app} and normalising yields~\eqref{eq:f_app}.

\subsection{Mixing Gibbs-Markov property}

We first establish the Gibbs-Markov property, and subsequently extend this to the mixing Gibbs-Markov property.

The Gibbs-Markov property is a regularity condition on expanding Markov maps, from which follows a body of useful limit theorems for ergodic sums (exponential decay of correlations, central limit theorem, stable and local limit theorems)~\cite{aaronson2001local,
heinrich1987rates}. The folded Gauss map is Gibbs-Markov as follows from the three conditions below~\cite{aaronson2001local}

\begin{enumerate}
\item \emph{Markov structure:} the Markov structure requires that the domain of $B$ can be partitioned into countably many intervals $I_p$, such that on each interval the action of $B$ is smooth, monotone, and invertible, and that the
image of each such interval coincides exactly with a union of such intervals, i.e. $B(I_p) = \bigcup_{q \in S_p} I_q$ for some set $S_p$, and cannot for example include part but not all of some $I_q$. For the folded Gauss map $B$ this holds in the strongest possible sense: each
$I_p$ maps homeomorphically onto the whole interval $[0,1/2]$, so
$B(I_p)$ is the union of all $I_p$. Explicitly, for integer
$k \geq 2$, 
\begin{equation}
    B(\talpha) =
    \begin{cases}
    \displaystyle\frac{1}{\talpha} - k
        & \quad \talpha \in I_{2k} := \bigl(\tfrac{2}{2k+1},\,\tfrac{1}{k}\bigr], \\[8pt]
    \displaystyle k+1 - \frac{1}{\talpha}
        & \quad \talpha \in I_{2k+1} := \bigl(\tfrac{1}{k+1},\,\tfrac{2}{2k+1}\bigr],
    \end{cases}
    \label{eq:map_B}
\end{equation}
where $\talpha$ lies in the interval $I_p$ with $p = \lfloor 2/\talpha \rfloor$. A direct check of~\eqref{eq:map_B} shows $ B(I_{p}) = [0,1/2]$ for all intervals $I_p$.

\item \emph{Big-image property.} No interval $I_p$ has a vanishingly small image. Trivially satisfied for $B$ as $B(I_p) = [0,1/2]$ for all intervals $I_p$.

\item \emph{Uniform regularity of the derivative under iteration.}
For a piecewise twice continuously differentiable Markov interval map satisfying (1) and (2)
above, together with uniform expansion $\inf|B'| > 1$, the
Gibbs-Markov distortion bound on iterates of $B$ is implied by
Adler's condition on $B$ itself~\cite{aaronson2001local},
\begin{equation}
    \sup_{\talpha \in (0,1/2)}\,
        \frac{|B''(\talpha)|}{|B'(\talpha)|^2} < \infty.
\end{equation}
For the folded Gauss map, $|B'(\talpha)| = 1/\talpha^2 \geq 4$ giving uniform expansion, and
\begin{equation}
    \frac{|B''(\talpha)|}{|B'(\talpha)|^2}= 2\talpha \leq 1
\end{equation}
giving Adler's condition. 
\end{enumerate}

The Gibbs-Markov property can be extended to include mixing on the further condition that

\begin{itemize}
    \item \emph{Mixing property:} For any two intervals $I_p, I_q$, there exists an integer $N$ such that
    \begin{equation}
        I_q \,\subset\, B^n(I_p) \pmod{0}
        \qquad \text{for every } n > N,
        \label{eq:mixing_def}
    \end{equation}
    that is, every interval is eventually covered by the forward iterates of every other interval~\cite{gouezel2010characterization}. For
    the folded Gauss map this holds trivially:
    $B(I_p) = [0,1/2]$ for every $I_p$, so $I_q \subset B(I_p)$ already at $n = 1$, and~\eqref{eq:mixing_def} is satisfied with $N = 0$
    uniformly in $p$ and $q$.
\end{itemize}

We note that the ergodicity and non-zero gap of $B$, verified in the previous sections, also follow from the mixing Gibbs-Markov property.

\subsection{The central-band super-step map $B(\talpha)$}
\label{app:Tc}

We briefly comment on how the results extend to the central band RG step. As given in Eq.~\eqref{eq:beta_scaling} of the main text, a single central step sends
$\talpha \to \talpha/(1-2\talpha)$. As noted, this generically results in negligible rescaling, so we
work with the \emph{super-step} of $r(\talpha) = \lfloor 1/(2\talpha)\rfloor$
consecutive central steps. The super-step is naturally written in terms of the rescaling $s(\talpha)$
\begin{equation}
    s(\talpha) = \frac{\talpha}{1-2\,r(\talpha)\,\talpha} \in [\tfrac12,\infty),
    \qquad r(\talpha) = \left\lfloor \frac{1}{2\talpha} \right\rfloor ,
    \label{eq:s_central}
\end{equation}
and subsequent folding back into the fundamental domain, resulting in 
\begin{equation}
    B(\talpha) = \min\big[b(\talpha), 1-b(\talpha)\big],
    \qquad b(\talpha) = s(\talpha) - \lfloor s(\talpha) \rfloor.
    \label{eq:map_Bc}
\end{equation}
it is straightforward to verify that a steady state distribution is given by
\begin{equation}
    [B f](\talpha) = f(\talpha) \qquad \text{for} \qquad f(\talpha) = \frac{2}{\log 3}\,\frac{1}{1 - \talpha^2},
\end{equation}
it follows that this is the unique steady state distribution by the gap property below.

\begin{itemize}

\item \textbf{Gibbs--Markov property:} The map $B$ can be recast in a form that is continuous and monotonic on a countable set of intervals
\begin{equation}
    B(\talpha) =
    \begin{cases}
    \displaystyle \frac{\talpha}{1-2 r \talpha} - k
        & \quad \talpha \in I_{2k,r} := \bigl(u_r(k),\,u_r(k+\tfrac12)\bigr], \\[8pt]
    \displaystyle k+1 - \frac{\talpha}{1-2 r \talpha}
        & \quad \talpha \in I_{2k+1,r} := \bigl(u_r(k+\tfrac12),\,u_r(k+1)\bigr],
    \end{cases}
    \qquad
    u_r(x) := \frac{x}{1+2rx},
    \label{eq:map_Bc2}
\end{equation}

We note that $B$ is Gibbs-Markov (i) each branch is full, $B(I_{n,r}) = [0,1/2]$, so the map is Markov (ii) The big-image condition is
satisfied and (iii) Adler's condition holds, specifically
\begin{equation}
    \frac{|B''(\talpha)|}{|B'(\talpha)|^2}
        = 4 r(\talpha)\,(1-2 \talpha r(\talpha))
        \le \frac{4r(\talpha)}{r(\talpha)+1} < 4 ,
    \label{eq:adler_central}
\end{equation}
where we have used that $1-2\talpha r(\talpha) \in \bigl(0,\, 1/(r+1)\bigr)$ across each interval $I_{n,r}$. The ratio is therefore bounded uniformly over all intervals.

\item \textbf{Chaoticity:}
Away from a measure-zero set the fold contributes a factor of unit modulus and
the derivative is set by stage one, $|B'(\talpha)| = (1-2r\talpha)^{-2}$.
Using ergodicity to average over $f$,
\begin{equation}
    \Lambda = \int_0^{1/2}\! d\talpha\, f(\talpha)\, \log|B'(\talpha)|
        = -2\!\int_0^{1/2}\! d\talpha\, f(\talpha)\, \log\!\big(1-2 r(\talpha)\talpha\big)
        = \frac{\pi^2}{2\log 3} ,
    \label{eq:lyapunov_central}
\end{equation}
so $B$ is chaotic. The per-super-step length rescaling is
$\log\lambda = \tfrac12 \log|B'|$, with mean
$\langle\log\lambda\rangle = \Lambda/2 = \pi^2/(4\log 3)$, giving
$\log(a_n/a_0)\sim \pi^2 n/(4\log 3)$ as in Eq.~\eqref{eq:aV_solution2}.

\item \textbf{Mixing Gibbs-Markov property:} The mixing property follows immediately from the structure
$B(I_{n,r}) = [0,1/2]$, which gives $I_{m,s} \subset B(I_{n,r})$
for every pair of intervals~\cite{gouezel2010characterization}.

\item \textbf{Ergodicity, and spectral gap:} Per Theorem~1.6 of Ref.~\cite{aaronson2001local}, the mixing Gibbs-Markov property established above implies that $B$ is gapped: the leading eigenvalue $1$ is simple, with corresponding eigenvector $f(\talpha)$, the remainder of the spectrum lies within a radius $r<1$. Hence $B$ has a positive spectral gap $\Delta > 0$. As this eigenvalue is unique, $B$ is further ergodic. We do not evaluate the gap $\Delta$ here, though one may do so by the monomial-basis method of Sec.~\ref{app:ergB}.
\end{itemize}

\section{Fluctuation analysis for error bars in Fig.~2}
\label{app:flucs}

In this appendix we consider the fluctuations on the sums defining $\log \tau$ and $\log \lambda$ (Eqs.~13 and~14)  given in the main text. The main results are the limiting scaling
\begin{equation}
    \begin{aligned}
        \log \lambda & = \sum_{k=0}^{n-1} \log \lambda(\talpha_k) \!\!\!\!&=&  K_\lambda n \left(1 + \frac{\eta_n}{\sqrt{n}} \right)
        \\
        \log \tau & = \frac{4 G}{\pi} \sum_{k=0}^{n-1} s(\talpha_k) \!\!\!\!&=& K_{\tau} n \log n  \left( 1 + \frac{\xi_n}{\log n} \right)
    \end{aligned}
\end{equation}
where $\eta_n$ is normally distributed, and $\xi_n$ is Landau distributed. 

We find that $\log \lambda$ is determined by the typical values of the sum, and hence $K_\lambda$ is the mean value of $\log \lambda(\talpha_k)$, whereas $\log \tau$ is determined by rare large values in the sum, and hence $K_\tau$ is set by the tail of the distribution $\rho(s)$ of the values of $s(\alpha_k)$
\begin{equation}
    K_\lambda = \int d \talpha f(\talpha) \log \lambda( \talpha), \qquad K_\tau = \frac{4 G}{\pi} \lim_{s \to \infty} s^2 \rho(s), \qquad \rho(s') = \int d \talpha f(\talpha) \delta(s'-s(\talpha)).
\end{equation}
the values of these objects are given in the main text. Combining these results we obtain the leading corrections to the asymptotic length-time rescaling 
\begin{equation}
    \frac{\log \tau}{\log \lambda} = \frac{K_\tau}{K_\lambda}\left(\log n + \xi_n+O(\log(n)/\sqrt{n})\right) = \zeta\Big(\log \log \lambda + c + z_n  +O(\log(n)/\sqrt{n})\Big)
    \label{eq:scaling}
\end{equation}
where $c + z_n = \xi_n - \log K_\lambda$. We find that $z_n$ is Landau distributed $z_n \sim \operatorname{Landau}(0,\tfrac{\pi}{2})$ (i.e. the distribution has characteristic function $\phi(t) = \exp( -it\log|t| - \tfrac{\pi}{2}|t|)$) and the coefficients are given by
\begin{equation}
    c = 
    \begin{cases}
        \displaystyle
        1 - \gamma_\mathrm{E} -\varphi \log \varphi - \log(\pi^2/12) & \qquad \text{(Lower band)}
        \\[6pt]
        \displaystyle
        1 - \gamma_\mathrm{E} - \log (3\pi^2/8) & \qquad \text{(Central band)}
    \end{cases}
    \label{eq:cvals}
\end{equation}
where $\gamma_\mathrm{E}$ is the Euler–Mascheroni constant. The theory lines in the main text Figs~2a and~2c then correspond to Eqs.~\eqref{eq:scaling} and~\eqref{eq:cvals} with $z_n$ replaced with the $\operatorname{Landau}(0,\tfrac{\pi}{2})$ distribution quartile values
\begin{equation}
    z_{1/4} = -0.204641\ldots,\qquad z_{1/2} = 1.35578\ldots, \qquad z_{3/4} = 4.45839\ldots.
\end{equation}


\subsection{A useful theorem}
The results of this section are obtained using a theorem due to Gou\"{e}zel~\cite{gouezel2010characterization} which concerns sums of the form 
\begin{equation}
    S_n = \sum_{k=0}^{n-1} h(\talpha_k), \qquad \talpha_k = B(\talpha_{k-1})
    \label{eq:Sn}
\end{equation}
for $B$ a mixing Gibbs Markov map. Specifically, theorem 1.5 of Ref.~\cite{gouezel2010characterization} states, under certain regularity conditions (discussed below) sums of the form~\eqref{eq:Sn} converge in distribution to L\'{e}vy alpha-stable distributions in the same manner as sums of correspondingly defined random variables. Specifically, there are two cases distinguished by whether the variance 
\begin{equation}
    \sigma_h^2 = \int d\talpha f(\talpha) (h(\talpha )- \mu_h)^2, \quad \text{where} \quad \mu_h = \int d\talpha f(\talpha) h(\talpha )
\end{equation}
is finite:
\begin{itemize}
    \item \textbf{Case 1, $\sigma_h^2 < \infty$}: Then the $S_n$ converge in distribution to a normal distribution. Specifically, the shifted and rescaled quantity
    \begin{equation}
        z_n=\frac{S_n - \mu n}{\sigma\sqrt{n}} \overset{d}{\longrightarrow} \mathrm{Normal}(0,1)
    \end{equation}
    is normally distributed with mean and variance given by
    \begin{equation}
        \mu = \int d \talpha f(\talpha)h(\talpha), \qquad \sigma^2   =\sum_{q=- \infty}^\infty c_{q}, \qquad c_q =\int d\talpha \, f(\talpha)\,(h(\talpha)-\mu)(h(B^{|q|}(\talpha))-\mu).
    \end{equation}
    The quantity $c_q$ has the natural interpretation as the covariance between $h(\talpha_k)$ and $h(\talpha_{k+q})$. We note that the normal distribution is the L\'{e}vy alpha-stable distribution for stability parameter $\alpha = 2$, connecting this result to the $\sigma_h^2 = \infty$ case.
    \item \textbf{Case 2, $\sigma_h^2 = \infty$:}. Then the $S_n$ converge in distribution to a L\'{e}vy alpha-stable distribution for stability parameter $0<\alpha \leq 2$. It then follows that
    \begin{equation}
        z_n = \begin{cases}
            \displaystyle
            \frac{S_n - n \delta}{\gamma n^{1/\alpha}} \quad & \text{for } \alpha \neq 1,
            \\[10pt]
            \displaystyle
            \frac{S_n - n \delta -\tfrac{2}{\pi}\beta \gamma n \log n }{\gamma n} & \text{for } \alpha = 1,
        \end{cases}
        \label{eq:shift_scaling}
    \end{equation}
    converges in distribution to a L\'{e}vy alpha-stable distribution
    \begin{equation}
        z_n \overset{d}{\longrightarrow} \operatorname{Stable}(\alpha,\beta,1,0)
        \label{eq:stable_dist}
    \end{equation}
    for some appropriate choice of $\alpha,\beta,\gamma,\delta$, and throughout we use the ``Nolan $S_0$'' convention for the stable distribution, so that $\operatorname{Stable}(\alpha,\beta,\gamma,\delta)$ has characteristic function $\phi$ given by
    \begin{equation}
        \log \phi(t; \alpha, \beta, \gamma, \delta) = i t \delta - |\gamma t|^\alpha \left (1 - i \beta \operatorname{sign}(t) \Phi(t) \right ), \qquad \Phi(t) = \begin{cases}
            \left ( 1 - |\gamma t|^{1 - \alpha} \right ) \tan \left (\tfrac{\pi \alpha}{2} \right )  & 
            \text{for }\alpha \neq 1 ,
            \\[1ex]
            - \frac{2}{\pi} \log|\gamma t| 
            & 
            \text{for }\alpha = 1.
        \end{cases}
    \end{equation}
    Ref.~\cite{gouezel2010characterization} further notes that parameters $\alpha,\beta,\gamma,\delta$ of the stable distribution may be computed via the precise relationship to the random variable
    \begin{equation}
        R_n = \sum_{k=1}^{n}h(\tbeta_k)
    \end{equation}
    where the $\tbeta_k$ are independent random variables which are drawn from the invariant measure $f$. Specifically the sums $S_n$ converge in distribution to the random variables $R_n$
    \begin{equation}
        S_n \overset{d}{\longrightarrow} R_n
    \end{equation}
    thus the $R_n$ converges to the same stable distribution~\eqref{eq:stable_dist} under the same rescaling~\eqref{eq:shift_scaling}
    allowing $\alpha,\beta,\gamma,\delta$ to be calculated directly from analysing the $R_n$ using standard techniques of generalised central limit theorem.
\end{itemize}

Theorem 1.5 of Ref.~\cite{gouezel2010characterization} holds subject to the convergence criterion
\begin{equation}
    \sum_{k} m(I_k)(D(h,I_k))^\eta< \infty
\end{equation}
for some $\eta \in (0,1]$. Here the $I_k$ are the intervals over which $B$ is smooth and monotone (see~\eqref{eq:map_B}), $m(I_k) = \int_{I_k} d \talpha f(\talpha)$ is the measure weight on each interval, and
\begin{equation}
    D(h,I) = \sup_{\substack{\talpha,\talpha'\in I\\\talpha \neq \talpha'}} \left| \frac{h(\talpha) - h(\talpha')}{\talpha - \talpha'}\right|
\end{equation}
is the best Lipschitz constant on the interval $I$. In the present context this condition is satisfied for any $h(\talpha)$ which has at most a finite order pole at the accumulation points of $B$ ($\talpha = 0$ for the lower band RG step, $\talpha = 1/(2p)$ for integer $p$ for the central band RG super step) and is smooth for all other $\talpha$. 

\subsection{Fluctuations on the length rescaling $\log \lambda$}

For the length rescaling $\log \lambda$ we have $\log \lambda = S_n$ with $h(\talpha) = \log \lambda(\talpha)$. As $h(\talpha)$ has only integrable divergences, then $\sigma_h^2$ is finite and Case 1 above applies. Consequently
\begin{equation}
    \log \lambda = n \int d \talpha f(\talpha) \log \lambda( \talpha) + z_{\lambda,n} \sigma_\lambda\sqrt{n} = nK_\lambda \left(1 + \frac{\eta_n}{\sqrt{n}} \right), \qquad \eta_n =  \frac{\sigma_\lambda}{K_\lambda} z_{\lambda,n}
\end{equation}
where the correction $z_{\lambda,n}$ are standard normally distributed variables, and $K_\lambda = \int d \talpha f(\talpha) \log \lambda( \talpha)$.

The fluctuations will be sub-leading to the $\log \tau$ fluctuations, and so it is not necessary or insightful to establish the value of $\sigma_\lambda$.

\subsection{Fluctuations on the time rescaling $\log \tau$}

For the time rescaling $\log \tau$, Case 2 detailed above applies, with $\log \tau = \tfrac{4 G}{\pi}S_n$, where per the theorem above, $S_n$ has the same distribution as 
\begin{equation}
    R_n = \sum_{k=1}^n s(\tbeta_k)
\end{equation}
for $\tbeta_k$ drawn iid from the invariant measure $f(\tbeta)$, where $s(\tbeta_k)$ and $f(\tbeta)$ are as in the main text
\begin{equation}
    s(\tbeta) = \begin{cases}
        \displaystyle
        \frac{1}{\tbeta} \qquad & \text{(lower band)}
        \\[5pt]
        \displaystyle
        \frac{\tbeta}{1 - 2 \tbeta \lfloor 1/(2 \tbeta)\rfloor}
        \qquad & \text{(central band)}
    \end{cases}
    \label{eq:app_s}
\end{equation}
\paragraph*{\textbf{Lower band RG}:} For the lower band, the characteristic function of $R_n$ is given by
\begin{equation}
    \begin{aligned}
        \log \phi(t) & = n \log \int_0^{1/2} d\tbeta f(\tbeta) e^{i t/\tbeta} 
        \\
        & =  n \log \int_{2}^{\infty} \frac{ds}{s^2} f(s^{-1}) e^{i s t}
        \\
        & = -\frac{\pi n}{2\log\varphi}|t|-\frac{in}{\log\varphi}t\log|t|+in s_*t+O\big(n t^{2}\log^{2}|t|\big)
    \end{aligned}
\end{equation}
where
\begin{equation}
    s_* = \frac{1-\gamma_E}{\log\varphi}-\varphi
\end{equation}
This corresponds to a stable distribution with $\alpha=1$, $\beta = 1$. By identifying the scales
\begin{equation}
    \begin{aligned}
        \gamma & = \frac{\pi}{2 \log \varphi}
        \\
        \delta & = \frac{1}{\log\varphi}\left(1 - \gamma_\mathrm{E} + \log\gamma\right) -\varphi
    \end{aligned}
\end{equation}
where $\gamma_\mathrm{E}$ is the Euler–Mascheroni constant, we find
\begin{equation}
    \frac{R_n - n \delta - \tfrac{2}{\pi}\gamma n \log n}{\gamma n} \sim \operatorname{Stable}(1,1,1,0) = \operatorname{Landau}(0,1)
\end{equation}
and hence $\log \tau$ converges in distribution
\begin{equation}
    \log \tau = \sum_{k=0}^{n-1} \log \tau_k =  \frac{4 G}{\pi}\left(\frac{2 \gamma }{\pi} n \log n + n \delta + z_{\tau,n} n \gamma \right) =K_\tau n (\log n + \xi_n) , \qquad  K_\tau =  \frac{4 G}{\pi \log \varphi}, \quad \xi_n = \log \varphi\left( \delta + z_{\tau,n} \gamma \right)
\end{equation}
with $z_{\tau,n}$ following a $\operatorname{Stable}(1,1,1,0)$ distribution. 

\paragraph*{\textbf{Central band RG}:} Repeating this analysis for the central band, with $s$ as in~\eqref{eq:app_s} yields a distribution of
\begin{equation}
    \rho(s') = \int d \talpha f(\talpha) \delta(s'-s(\talpha)) = \frac{1}{\log 3} \frac{1}{s'(1+s')}, \quad s'\in[\tfrac12,\infty)
\end{equation}
The characteristic function of $R_n$ is then
\begin{equation}
    \begin{aligned}
        \log\phi(t) &= n\log\int_{0}^{1/2}d\tbeta\, f(\tbeta)\,e^{i t\, s(\tbeta)}
        = n\log\int_{1/2}^{\infty} ds\, \rho(s)\, e^{i s t}
        \\
        &= -\frac{\pi n}{2\log 3}|t|-\frac{in}{\log 3}\,t\log|t|+i n\, s_*\, t
        +O\big(n\, t^{2}\log^{2}|t|\big),
    \end{aligned}
\end{equation}
where
\begin{equation}
    s_* = \frac{1-\gamma_\mathrm{E}+\log 2}{\log 3}-1 .
\end{equation}
This again corresponds to a stable distribution with $\alpha=1$, $\beta=1$. Identifying
the scales
\begin{equation}
    \begin{aligned}
        \gamma &= \frac{\pi}{2\log 3}
        \\
        \delta &= \frac{1}{\log 3}\left(1-\gamma_\mathrm{E}+\log(2\gamma)\right)-1 ,
    \end{aligned}
\end{equation}
we find $\big(R_n-n\delta-\tfrac{2}{\pi}\gamma n\log n\big)/(\gamma n)\sim\operatorname{Stable}(1,1,1,0)$,
and hence
\begin{equation}
    \log\tau = \sum_{k=0}^{n-1}\log\tau_k
    = \frac{4G}{\pi}\left(\frac{2\gamma}{\pi}n\log n+n\delta+z_{\tau,n}n\gamma\right)
    = K_\tau n\,(\log n+\xi_n),
    \qquad K_\tau = \frac{4G}{\pi\log 3}, \quad \xi_n=\log 3\,(\delta+z_{\tau,n}\gamma),
\end{equation}
with $z_{\tau,n}$ following a $\operatorname{Stable}(1,1,1,0)$ distribution.

Having obtained the two scalings, we combine them to recover the
length--time relation quoted in the introduction. Writing $n$ in terms of the
physical scale via $\log\log\lambda = \log n + \log K_\lambda + O(1/\sqrt{n})$
and forming the ratio,
\begin{equation}
    \frac{\log\tau}{\log\lambda}
    = \zeta\Big(\log\log\lambda + c + z_n + O\!\big(\tfrac{\log n}{\sqrt{n}}\big)\Big),
    \qquad \zeta = \frac{K_\tau}{K_\lambda},
    \quad c + z_n = \xi_n - \log K_\lambda,
\end{equation}
which is Eq.~\eqref{eq:scaling}. The deterministic offset $c$ takes the
band-dependent values in Eq.~\eqref{eq:cvals}, while the fluctuation
$z_n \sim \operatorname{Landau}(0,\pi/2)$ is band-independent.

\end{widetext}

\end{document}